\documentclass[traditabstract, longauth]{aa} 
\usepackage{txfonts}
\usepackage{graphicx}
\usepackage{color}
\usepackage{natbib}
\bibpunct{(}{)}{;}{a}{}{,}
\def\gapprox{{_>\atop{^\sim}}}
\def\lapprox{{_<\atop{^\sim}}}

\def\cmmd{\rm {cm^{-3}}}
\def\cmmt{\rm {cm^{-2}}}

\def\s-1{\rm {s^{-1}}}

\def\HC3N{HC$_3$N}

\def\kms{\hbox{${\rm km\,s}^{-1}$}}
\def\msun{M$_{\odot}$}
\def\lsun{L$_{\odot}$}

\newcommand{\asec}{\mbox{$''$}}

\usepackage{url}

\begin{document}
 \title{The hidden heart of the luminous infrared galaxy IC~860 ${\rm I}$.\\
{\it  A molecular inflow feeding opaque, extreme nuclear activity}\thanks{Based on observations
carried out with the ALMA Interferometer. ALMA is a partnership of ESO (representing its member states),
NSF (USA) and NINS (Japan), together with NRC (Canada) and NSC and ASIAA (Taiwan),
in cooperation with the Republic of Chile. The Joint ALMA Observatory is operated by ESO, AUI/NRAO and NAOJ.}}

\author{S. Aalto
          \inst{1}
          \and
        S. Muller\inst{1}
          \and
	  S. K\"onig\inst{1}
	   \and
	  N. Falstad\inst{1}
          \and
          J. Mangum\inst{2}
          \and
 	K.  Sakamoto\inst{3}
          \and
 	G. C. Privon\inst{4}
  	  \and
          J. Gallagher\inst{5}
	\and
         F. Combes\inst{6}
         \and
	  S. Garc\'ia-Burillo\inst{7}
         \and
	 S. Mart\'in\inst{8}
	 \and
 	S. Viti\inst{9}
        \and	
          P. van der Werf\inst{10}
        	\and
 	A. S. Evans\inst{11}
	\and
 	J.H. Black\inst{1}
		\and
 	E. Varenius\inst{12}
	\and
 	R. Beswick\inst{12}
	\and
	G. Fuller\inst{12}
	\and
	C. Henkel\inst{13, 17}
	\and
	 K.  Kohno\inst{14}
	\and
 	K. Alatalo\inst{15}	
	\and
       S. M\"uhle\inst{16}
             }

 \institute{Department of Space, Earth and Environment, Chalmers University of Technology, Onsala Observatory,
              SE-439 92 Onsala, Sweden\\
              \email{saalto@chalmers.se}
\and  National Radio Astronomy Observatory, 520 Edgemont Road, Charlottesville, VA 22903-2475, USA           
\and  Institute of Astronomy and Astrophysics, Academia Sinica, PO Box 23-141, 10617 Taipei, Taiwan 
\and Department of Astronomy, University of Florida, PO Box 112055, USA
\and  Department of Astronomy, University of Wisconsin-Madison, 5534 Sterling, 475 North Charter Street, Madison WI 53706, USA
\and Observatoire de Paris, LERMA (CNRS:UMR8112), 61 Av. de l'Observatoire, 75014 Paris, France 
\and Observatorio Astron\'omico Nacional (OAN)-Observatorio de Madrid, Alfonso XII 3, 28014-Madrid, Spain
\and European Southern Observatory, Alonso de Cordova 3107, Vitacura, Santiago, Chile
\and Department of Physics and Astronomy, UCL, Gower St., London, WC1E 6BT, UK 
\and Leiden Observatory, Leiden University, 2300 RA, Leiden, The Netherlands
\and University of Virginia, Charlottesville, VA 22904, USA, NRAO, 520 Edgemont Road, Charlottesville, VA 22903, USA
\and Jodrell Bank Centre for Astrophysics, School of Physics \& Astronomy, University of Manchester, Oxford Road, Manchester M13 9PL, UK
\and Max-Planck-Institut f\"ur Radioastronomie, Auf dem H\"ugel 69, 53121 Bonn, Germany
\and Institute of Astronomy, The University of Tokyo, Osawa, Mitaka, Tokyo 181-0015, Japan
\and Space Telescope Science Institute, 3700 San Martin Drive, Baltimore, MD, 21218, USA
\and Argelander-Institut f\"ur Astronomie, Auf dem H\"ugel 71, 53121 Bonn, Germany
\and Astron. Dept. King Abdulaziz University, P.O. Box 80203, Heddah 21589, Saudi Arabia
 }

   \date{Received xx; accepted xx}

  \abstract{High-resolution (0.\asec 03 to 0.\asec 09 (9 to 26 pc))  ALMA (100 to 350 GHz ($\lambda$ 3 to 0.8~mm)) and (0.\asec 04 (11 pc))  VLA 45 GHz  measurements have been
  used to image continuum and  spectral line emission from the inner (100 pc) region of the nearby infrared luminous galaxy IC~860.   We detect compact ($r \sim $10 pc), luminous,
 3 to 0.8 mm continuum emission in the core of IC~860, with brightness temperatures $T_{\rm B}>160$ K.  The 45 GHz continuum is equally compact but significantly fainter in flux.
We suggest that the  3 to 0.8~mm continuum emerges from hot dust with radius $r$$\sim$8 pc and temperature $T_{\rm d}$$\sim$280 K, and that it is opaque at millimetre wavelengths, implying a very large 
 H$_2$ column density $N$(H$_2$)$\gapprox$$10^{26}$ $\cmmt$.  
 
Vibrationally excited lines of HCN  $\nu_2$=1f J=4--3 and 3--2 (HCN-VIB)  are seen in emission and spatially resolved on scales of 40-50 pc. The line-to-continuum ratio drops towards the
inner $r$=4 pc, resulting in a ring-like morphology. This may be due to high opacities and matching HCN-VIB excitation- and continuum  temperatures.  The HCN-VIB emission reveals a north--south nuclear velocity
gradient with projected rotation velocities of $v$=100  \kms\ at $r$=10 pc. The brightest emission is oriented perpendicular to the velocity gradient, with a peak HCN-VIB 3--2 $T_{\rm B}$ of 115 K (above the continuum).

Vibrational ground-state lines of HCN 3--2 and 4--3, HC$^{15}$N 4--3, HCO$^+$ 3--2 and 4--3, and CS 7--6 show complex line absorption and emission features towards the
dusty nucleus. Redshifted, reversed P-Cygni profiles are seen for HCN and HCO$^+$   consistent with gas inflow with $v_{\rm in} \lapprox 50$ \kms. Foreground absorption structures 
outline the flow, and can be traced from the north-east into the nucleus. In contrast, CS 7--6 has blueshifted line profiles with line wings extending out to -180 \kms. 
We suggest that a dense and slow outflow is hidden behind a foreground layer of obscuring, inflowing gas. 

The centre of IC~860 is in a phase of rapid evolution where an inflow is building up a massive nuclear column density of gas and dust that feeds star formation and/or AGN activity.
The slow, dense outflow may be signaling the onset of feedback. The inner, $r$=10 pc, IR luminosity may be powered by an AGN or a compact starburst, which then would require top-heavy initial mass function. }
 

\titlerunning{The dark heart of the luminous infrared galaxy IC~860}

 \maketitle


\section{Introduction}

Luminous  ($L_{\rm IR}$=$10^{11}-10^{12}$ \lsun) and ultraluminous ($L_{\rm IR}\gapprox 10^{12}$ \lsun) infrared galaxies (U/LIRGS) are powered by either  bursts of 
star formation or accreting supermassive black holes (SMBHs), and are important to our understanding of galaxy growth throughout the
Universe \citep[e.g.][]{elbaz03,sanders96}.  How the accreting SMBHs (active galactic nuclei (AGN)) evolve together with their host galaxies is one of the fundamental questions in galaxy evolution.
Some U/LIRGs harbour very deeply embedded nuclei that appear to go through a stage of rapid growth 
\citep[e.g.][]{sakamoto08, spoon13, gonzalez12, aalto15b}. Studying these dark nuclei is essential for a complete AGN and starburst census, for constraining orientation-based
unification models, and to probe the onset of feedback processes in dusty galaxies 
 \citep[e.g.][]{brightman12, merloni14}.  Of prime importance is the nature of the buried source(s) that powers the luminosity of the enshrouded nucleus.

Recent centimetre (cm) and millimetre (mm) observations of exceptionally luminous emission (and absorption at $\lambda$=6 cm) from vibrationally
excited HCN have been reported in a number of U/LIRGs \citep{salter08, sakamoto10, imanishi13, aalto15a, aalto15b, martin16}. The emission emerges from
compact ($r<15-75$ pc), hot ($T>$100 K), and opaque ($N$(H$_2$)$>10^{24}$ $\cmmt$) regions centred on the nuclei. 
In contrast, we find for several cases that the vibrational ground-state\footnote{From now on we refer to the "vibrational ground state"  as the "ground state".} HCN and HCO$^+$ 3--2, 
4--3 emission lines suffer from continuum- and self-absorption towards the inner 100 pc of these galaxies.
Their central (and sometimes also global) line profiles become double-peaked since photons at the line centre become absorbed by foreground, cooler gas. 
Self-absorption can be caused by a temperature gradient and large line-of-sight column densities \citep{aalto15b}.  Therefore, lines like ground-state HCN and HCO$^+$ (standard tracers
of dense ($n>10^4$ $\cmmd$) gas) will show complex line profiles and their absorption spectra may not probe all the way through to the inner region of the galaxy nucleus. 
Lines of vibrationally excited molecules such as HCN will reach further inside the opaque layers. Because they often couple to high-surface-brightness radiation fields, the
line emission from vibrationally excited molecules can probe the buried dynamical mass and also hidden high-surface brightness IR nuclei - attenuated at their intrinsic wavelengths. 

These { compact obscured nuclei (CONs)} may also be missed in surveys using IR spectral energy distributions (SEDs) to identify galaxies with hot, dense nuclei. Galaxies with CONs may 
masquerade as normal starbursts since their SEDs can appear relatively cool. The CONs have luminous dust continuum that can be opaque down to long wavelengths \citep[e.g.][]{sakamoto08, sakamoto13, costagliola13, sakamoto17}. The emission from the hot nucleus becomes attenuated by the outer layers of dust, thereby shifting the energy density to
longer wavelengths.  However,  interior temperatures may be very high (exceeding 200 K) and the nuclei may be partly "self-heating" (depending on geometry)
from the trapping of emission, (i.e. a form of  "greenhouse effect" for galaxy nuclei - a process that has been suggested to occur on smaller scales for Galactic hot cores \citep{kaufman98,rolffs11b}.  

Vibrationally excited HCN (HCN-VIB) is excited by mid-infrared (MIR) 14 $\mu$m continuum up to energy levels $T_{\rm E}>1000$ K,  in contrast to the ground-state HCN and HCO$^+$ $J$=3--2 and 4--3,
lines which have $T_{\rm E}$ of 20-40 K. The vibrational ladder has transitions in the mm and sub-mm band that can be observed with ALMA  \citep{ziurys86,sakamoto10,aalto15b}. 
The HCN-VIB lines require intense MIR emission to be excited with intrinsic surface brightness $\Sigma_{\rm MIR} > 5 \times 10^{13}$ \lsun\, kpc$^{-2}$.  When H$_2$ column densities exceed
$N$(H$_2$)$>10^{24}$ $\cmmt$,  X-rays and MIR are strongly attenuated, while the HCN-VIB lines require large column densities
to be detectable \citep{aalto15b}.  So far we have found that the HCN-VIB line emission in CONs is too luminous with respect to the $L$(IR) to represent a normal cool mode of star
formation \citep{aalto15b}.   The intense HCN-VIB emission may instead be emerging from Compton-thick (CT) AGNs powered by accreting SMBHs or from an embedded, compact burst of 
star formation: the hot ($T>200$ K), opaque starburst  \citep{andrews11}.  The most rapidly evolving SMBHs are expected to be deeply embedded and the HCN-VIB lines allow us to probe the
most obscured phase of nuclear accretion. Studies suggest that up to 50\% of  low-luminosity AGNs may be obscured in X-rays and the MIR \citep[][e.g.]{Lusso13}, and only a small fraction of them have been
identified to date. In addition, a recent study suggests that the CONs may represent an early and/or compact stage in the onset of nuclear, dusty feedback \citep{falstad19}.

\medskip
\noindent
The LIRG IC~860  ($D$=59 Mpc, log $L_{\rm IR}$=11.17 \lsun, 1\asec=286 pc; \citet{sanders03})  is one of the most nearby CONs with very luminous and compact
HCN-VIB emission \citep{aalto15b} and is an ideal object with which to study the distribution and dynamics of the HCN-VIB emission and the structure of the nuclear continuum. 
IC~860 is a barred galaxy with a post-starburst optical spectrum \citep{alatalo16} and with {\ion H i} and OH absorption towards the centre \citep{schmelz86, kazes88}. Mid-infrared silicate absorption
and a low [C II] 157.7 $\mu$m-to-$L_{\rm FIR}$ ratio suggest a warm, compact, and obscured inner region of IC~860  \citep[e.g.][]{spoon07, diaz13}. The classification of IC~860 as a starburst or AGN is strongly aggravated
by the layers of dust \citep[e.g.][]{alonso06}.

\medskip
\noindent
In this paper we present high-resolution Atacama Large Millimeter Array (ALMA) band 3, 6, and 7 observations
of the 3 to 0.8~mm continuum and $J=3\to 2$, $4\to 3$, $\nu$=0  and $\nu_2$=1f HCN, $J=3\to 2$, $4\to 3$ HCO$^+$, $J=4\to 3$ HC$^{15}$N, and $J=7\to 6$ CS in IC~860. 
We also present high-resolution Very Large Array (VLA)  Q-band (45 GHz) continuum observations. 


\section{Observations}
\label{s:obs}

\subsection{ALMA observations}
For the ALMA observations the phase centre was set to 
$\alpha$=13:15:03.5088  and $\delta$=$+$24:37:07.788 (J2000). All data
were calibrated  within the CASA reduction package. The visibility set was then imported into the AIPS
package for further imaging. A journal of the ALMA observations is presented in Table~\ref{t:obs_ALMA},
and Table~\ref{tab:obs_spectral} lists the ALMA spectral set up.  We use
the procedure {\it checksource}\footnote{See e.g. ALMA cycle 6 Technical Handbook: 
https://almascience.eso.org/documents-and-tools/cycle6/alma-technical-handbook} to determine the accuracy of the astrometry.  
The quasars used for bandpass, complex gain, flux calibration and check source are listed in Table~\ref{t:obs_ALMA}.

For ALMA band 3, the  astrometry is offset by 0.2$\times$ beam and for band 6 the offset is 0.5$\times$beam.  For band 7 the offset
is estimated to 0.6-0.8 times the beam. For band 7 the source we intended to use for {\it check source} was very weak rendering the comparison difficult and furthermore the
dataset did not meet the proposal RMS requirements. Flux calibration errors are 5-10\% for band 3, and 20\% for band 6. For band 7
we estimate a flux accuracy of 50 to 70\% due to issues with the flux calibrator.

\begin{table*}[h]
\caption{Journal of the ALMA observations.}
\label{t:obs_ALMA}
\begin{center} \begin{tabular}{ccccccccccc}
    \hline
Band & Date of      & N$_{\rm ant}$  & PWV $^{(c)}$ & t$_{\rm on}$ $^{(d)}$ & B$_{\rm min}$ / B$_{\rm max}$ $^{(e)}$ & Bandpass        & Flux       & Gain       & Check \\
     & observations & $^{(b)}$      & (mm)         & (min)              & (m / km)                          & calibrator     & calibrator  & calibrator & source & \\
    \hline

B3 $^{a1}$ & 2017 Sep 19 & 44 & $\sim 1$ & 40 & 41 / 12.1 & J1337$-$1257 & J1337$-$1257 & J1303+2433  & J1314+2348 \\
\hline
B6 $^{a1}$ & 2017 Nov 12 & 45 & $\sim 1$ & 30 & 113 / 13.9 &  J1256$-$0547 & J1229+0203  & J1327+2210 & J1314+2348 & \\
\hline
B7 $^{a2}$ & 2015 Nov 24 & 36 & $\sim 1.5$ & 25 & 16 / 12.5 & J1256$-$0547 & J1229+0203 & J1303+2433 & J1321+2216 & \\
   & 2015 Dec 05 & 39 & $\sim 1.6$ & 40 & 16 / 6.1 & J1256$-$0547 & J1229+0203 & J1303+2433 & J1321+2216 & \\

\hline
\end{tabular} \end{center}
 \mbox{\,} \vskip -.5cm
 $(a1)$ ALMA project number 2016.1.00800.S; $(a2)$ ALMA project number 2015.1.00823.S;
 $(b)$ Number of 12\,m-antennas in the array;
 $(c)$ Amount of precipitable water vapor in the atmosphere;
 $(d)$ On-source time;
 $(e)$ Respectively minimum and maximum projected baseline.  Largest recoverable scale: B7 0.\asec 25, B6 0.\asec 29, B3 0.\asec 66.
\end{table*}

\begin{table}[h]
\caption{The spectral setups of the ALMA observations.}
\label{tab:obs_spectral}
\begin{tabular}{ccl}
    \hline
    Sky frequency  & Bandwidth  & Main line \\
          (GHz)        & (GHz)    &  \\
    \hline
    96.7  & 1.875 & CS 2--1\\
       98.1  & 2.0   &  Cont. \\
       108.2 & 1.875 &  HC$_3$N \\
       110.1 & 1.875 &  $^{13}$CO 1--0 \\
    \hline
    246.3 & 2.0   & Cont.\\
    248.2 & 1.875 &  CH$_2$NH \\
    262.5 & 1.875 &  HCN 3--2\\
    264.1 & 1.875 &  HCO$^+$ 3--2\\
    \hline
    338.1 & 1.875 &  Cont.\\
       339.9 & 1.875 &  Cont. \\
       350.1 & 1.875 & HCN 4--3\\
       351.9 & 1.875 &  HCO$^+$ 4--3\\
    \hline
\end{tabular} 

\end{table}

\subsubsection{ALMA band 7}
\label{s:obs_ALMA_B7}
Observations were carried out with 36 and 39 antennas in the array on November 24 and December 5, 2015, 
for $\sim$20 minutes on-source ($\sim$40 minutes in total) and with reasonable atmospheric conditions
(system temperature: average $T_{\rm sys}\simeq$300 K)). 

The correlator was set up to cover two bands of 1.875~GHz in spectral mode, one centred at a
frequency of $\sim$350~GHz to cover HCO$^+$  $J$=4--3 and the vibrationally excited HCN  $J$=4--3
$\nu_2$=1f line, and the other to cover continuum at frequencies 338 to 340~GHz, which also contains CS $J$=7--6.

The synthesized beam is $0.\asec 036 \times 0.\asec 026$ with Briggs weighting (robust parameter set to 0.5). 
The resulting data have a sensitivity of 1.4~mJy per beam in a 20~\kms\ (24~MHz) channel width. For
natural weighting the synthesized beam is $0.\asec 047 \times 0.\asec 039$ and the sensitivity 1.2~mJy per beam
in the 20~\kms\ channel width.

\subsubsection{ALMA band 6}
\label{s:obs_ALMA_B6}
Observations  were carried out with 45 antennas in the array on November 12, 2017,
for $\sim$30 minutes on-source ($\sim$56 minutes in total) and with reasonable atmospheric
conditions (system temperature: average $T_{\rm sys}\simeq$100 K). 

The correlator was set up to cover three bands of 1.875~GHz in spectral mode, one centred at a
frequency of $\sim$264.1~GHz to cover HCO$^+$  $J$=3--2 and the vibrationally excited HCN  $J$=3--2
$\nu_2$=1f line (in the lower side band), and one centred at 262.5~GHz to cover HCN $J$=3--2.
The third was  centred at a frequency of $\sim$248.1~GHz to cover CH$_2$NH 6(0,6)-5(1,5). One
continuum band was centred on 246.3 GHz.

The synthesized beam is $0.\asec 05 \times 0.\asec 02$ with Briggs weighting (parameter robust set to 0.5). 
The resulting data have a sensitivity of 0.35~mJy per beam in a 20~\kms\ (18~MHz) channel width.

\subsubsection{ALMA band 3}
\label{s:obs_ALMA_B3}
Observations were carried out with 44 antennas in the array on September 19, 2017, 
for $\sim$40 minutes on-source ($\sim$67  minutes in total) and with reasonable atmospheric conditions
(system temperature: average $T_{\rm sys}\simeq$65 K). 

The correlator was set up to cover three bands of 1.875~GHz in spectral mode, one centred at a
frequency of $\sim$108.19~GHz to cover HC$_3$N  v7=1 $J$=12--11, one
centred at 110.1~GHz to cover $^{13}$CO $J$=1--0,  and the third centred at 96.72 GHz on
CS $J$=2--1. A continuum band was centred on 98.1 GHz

The synthesized beam is $0.\asec 1 \times 0.\asec 07$ with Briggs weighting (parameter robust set to 0.5). 
The resulting data have a sensitivity of 0.15~mJy per beam in a 20~\kms\ (7~MHz) channel width.

\subsection{Observations with the VLA}

Two Q-band basebands covering 1024\,MHz were measured toward IC~860 on July 2, 2015,
for a total of 70 minutes on-source with the VLA in its A configuration.  These
measurements were amplitude, bandpass, and phase calibrated with
observations of $1331+305$ (3C\,286;  amplitude and bandpass; flux
density = 1.4\,Jy) and J$1327+2210$ (Flux density =
$0.5028\pm0.001$\,Jy beam$^{-1}$ at 45.1\,GHz; spectral index =
$-0.28\pm0.099$), respectively, using standard techniques.

The correlator with the two 1024\,MHz basebands covered
sky frequencies from 43638 to 44662\,MHz and 45588 to 46612\,MHz.
Each baseband was subdivided sequentially into eight dual-polarization
spectral windows each of which contained 128  channels, each 1\,MHz in width.  
Two additional 128\,MHz spectral windows were configured to measure
the X-band (9~GHz) continuum emission for pointing recalibration.  
The original aim was to target several spectral lines, but there were no line
detections. The 7.5 \kms\ wide spectral channel RMS for the
spectral window image cubes was in the range 0.9 to 1.3\,mJy beam$^{-1}$.
The theoretical RMS is predicted to be $\sim 1.26$\,mJy beam$^{-1}$.
The synthesized beam is $0.\asec 047 \times 0.041$ with Briggs
weighting (parameter robust set to 0.5). 

Calibration of these observations resulted in $\sim40$\% of the
IC~860 measurements being flagged mainly due to antenna-not-on-source
errors.  This calibration also resulted in two of the spectral windows
being totally flagged, which resulted in a total detection bandwidth
of 1792\,MHz.  


\section{Results}

\subsection{Continuum}
\label{s:cont}

\begin{figure*}
\resizebox{18.2cm}{!}{\includegraphics[angle=0]{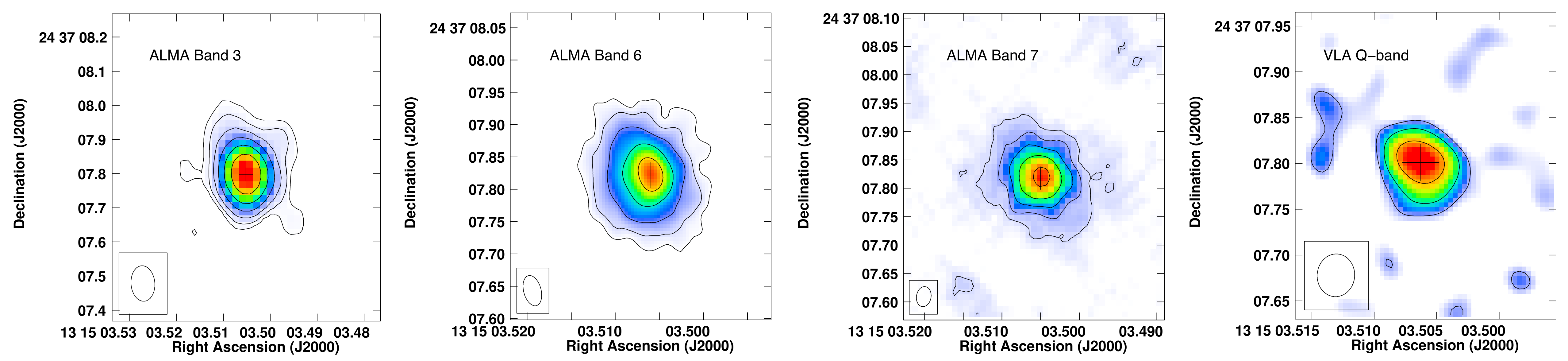}}
\caption{\label{f:cont} Band 3 continuum: Colour scale is 0.054 to 3 mJy; contours $0.018 \times$(3,6,12,24,48,96) mJy. Band 6 continuum:  Colour scale is 0.57 to 15 mJy; contours $0.19 \times$(3,6,12,24,48) mJy
Band 7: Colour scale is $>$0.039 to $>$8 mJy; contours $>0.13 \times$(3,6,12,24,48) mJy; Band~Q:  Colour scale is 0.05 to 0.5 mJy; contours $0.05 \times$ (3,6,12,24,48) mJy. The size of the synthesized beam is indicated in 
the lower left corner of each panel. The crosses mark the continuum peaks at each band (see Sec.~\ref{s:cont}).}
\end{figure*}

We merged line-free channels to produce continuum images from 45 up to 360 GHz.
We detect continuum at all observed wavelengths and in Fig.~\ref{f:cont} we present the continuum images.  Continuum fluxes, 
FWHM source sizes, fitted brightness temperatures ($T_{\rm B}$), and continuum positions can be found in Table~\ref{t:cont}. 
For a distance $D$ of 59 Mpc, a size scale of 10 pc corresponds to 0.\asec 035.

\subsubsection{ALMA continuum}
We detect luminous ALMA band 7 (339-350 GHz), band 6 (239-265 GHz), and band 3 (96-110 GHz) continuum emission.

\smallskip
\noindent
{\it Band 7:} \,  The peak flux is $>$7.0 mJy (356 GHz) and the integrated flux is $>$42 mJy (RMS $>$0.13 mJy). The  fluxes are lower limits due
to issues with the calibration (see Sect.~\ref{s:obs_ALMA_B7}).
Continuum structure is resolved with a source size $\theta$=0.\asec067 $\times$ 0.\asec064. Lower-surface-brightness emission can be found on scales of 200-300 mas.  

\smallskip
\noindent
{\it Band 6:} \,  The peak flux (265 GHz) is 13.4 mJy (RMS 0.2 mJy) and the integrated flux is 49 mJy. The continuum is resolved with a source size
$\theta$=0.\asec061 $\times$ 0.\asec055 and elongated with a position angle PA=22$^{\circ}$
with an inner compact structure and an extension to the east.   Lower-surface-brightness emission can be found on scales of 0.\asec2-0.\asec03.

\smallskip
\noindent
{\it Band 3:} \, The peak flux is 3.3 mJy (100 GHz) (RMS 0.04 mJy) and the integrated flux is 5.0 mJy .  The continuum structure has a FWHM source size
$\theta$ of 0.\asec066 $\times$ 0.\asec057. The shape of the continuum is extended to the north with PA=16$^{\circ}$, and with an extra extension to the
northwest. Lower-surface-brightness emission can be found on scales of 0.\asec2 - 0.\asec3. 

\subsubsection{The VLA continuum}

The Q-band continuum structure is resolved with a source size $\theta$ of 0.\asec059 $\times$ 0.\asec037. The peak flux is 0.54 mJy beam$^{-1}$ (45 GHz)
(RMS 0.01 mJy) and the integrated flux is 1.16 mJy. The shape of the continuum is somewhat elongated with PA=56$^{\circ}$. 


\begin{table*}[tbh]
\caption{\label{t:cont} Continuum fluxes and source sizes$^{\dag}$.}
\begin{tabular}{lcccccccc}
\hline
\hline
\\[0.1mm]
Frequency & Peak & Integrated & $\theta_{\rm source}$ & $\theta_{\rm beam}$ & $T_{\rm B}$ & R.A. & Dec. &  Positional\\
 &&&&&&&&  uncertainty \\
   (GHz)       & (mJy\,beam$^{-1}$) & (mJy)  & (mas) & (mas) & (K) & J(2000) & J(2000) & (mas) \\
\hline
\\[0.1mm]
45 &  $0.54 \pm 0.01$ & $1.16 \pm 0.02$ & 59 $\times$ 37  &  47 $\times$ 41 & 318 & 13:15:03.506 & +24:37:07.800 & 17 \\ 
100  &  $3.3 \pm 0.04$ & $5.04 \pm 0.08$ & 66 $\times$ 57  &  100 $\times$ 70 & 162 & 13:15:03.506 & +24:37:07.808 & 17\\ 
245$^a$   &  $13.3 \pm 0.2$ & $42 \pm 1$ & 59 $\times$ 54  &  50 $\times$ 20 & 266 & 13:15:03.506 & +24:37:07.823 & 20 \\
265$^b$   &  $13.4 \pm 0.2$ & $49 \pm 5$ & 61 $\times$ 55  &   59 $\times$ 37 & 250 & 13:15:03.506 & +24:37:07.823 & 20\\
356           &  7.4:          & 50:         & 67 $\times$ 60  &  36 $\times$ 26 & $>$120 & 13:15:03.505 & +24:37:07.818 & 25\\

\hline
\end{tabular} 
\\
\newline
\\
\begin{minipage}[h]{0.9\hsize}
$^{\dag}$Continuum levels were determined through a zeroth-order fit to line-free channels in the uv-plane for
Q-band,  and in the image plane for the ALMA continuum. Source sizes (diameters) are FWHM two-dimensional Gaussian fits given in 
(mas)=milli arcseconds=0.\asec001. Given flux errors are RMS errors only.  Flux calibration accuracy is 5-10\% for band 3 and Q-band, and 20\% for band 6. For band 7 the accuracy is
only a factor of two.   $T_{\rm B}$ is the Rayleigh-Jeans temperature over the 2D Gaussian source size. For a distance $D$ of 59 Mpc, a size scale of 10 pc corresponds to 0.\asec 035.
$^a$Lower side band (LSB) of the band 6 observations.
$^b$Upper side band (USB) of the band 6 observations. Errors are higher than for the LSB due to more line contamination.
\end{minipage}
\end{table*}

\subsection{Spectral lines}

\begin{figure}
\resizebox{6cm}{!}{\includegraphics[angle=0]{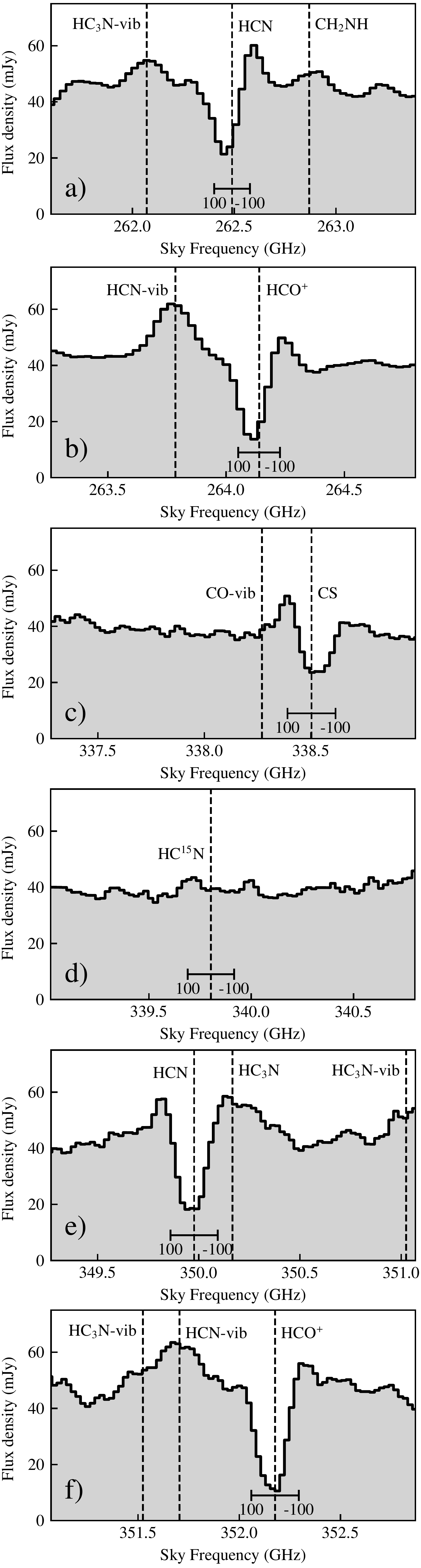}}
\caption{\label{f:all_spectra}  Band 6 and band 7 average spectra of the inner 0.\asec 1 of the central region. a) HCN 3--2, HC$_3$N $\nu_7$=2, CH$_2$NH; b)  HCO$^+$ 3--2, and HCN-VIB 3--2; c) CS 7--6 with 
a vibrational line of CO also marked; d) HC$^{15}$N 4--3; e) HCN 4--3, HC$_3$N 39--38,  HC$_3$N $\nu_7$=1; f) HCO$^+$ 4--3, HCN-VIB 4--3, and HC$_3$N $\nu_7$=2.
 The dashed lines mark $v_{\rm sys}$ ($cz$=3880 \kms) of the various species. We also indicate $\pm$100 \kms\ velocities for several of the species.  }
\end{figure}

In Fig.~\ref{f:all_spectra} we show ALMA band 6 and 7 spectra, averaged over the inner 0.\asec 1. We detect (for example)  prominent lines of HCO$^+$, and HCN 3--2 ground state $\nu$=0 and
vibrationally excited $\nu_2$=1, and HC$_3$N (vibrationally excited $\nu_7$=2). In band 7 we detect lines of HCO$^+$, HC$^{15}$N, and HCN 4--3 (ground
state $\nu$=0 and vibrationally excited $\nu_2$=1), HC$_3$N (both ground state and vibrationally excited). It is clear from the spectra that on 0.\asec 1 size scales many of the ground state 
lines are dominated by absorption rather than emission.

In the following sections we first present the results of the vibrationally excited HCN, followed by the ground-state lines. Here, we  adopt $cz$=3880 $\pm$ 20 \kms\ ($z$=0.01294) for systemic velocity ($v_{\rm sys}$)
from \citet{aalto15b}; systemic velocity is further discussed in Sect.~\ref{s:hcn-vib_dyn}.

We present integrated intensity (moment 0), velocity field (moment 1), and dispersion (moment 2) maps for the part of the lines (or parts of lines) that are emitting above the continuum.
For all lines we (conservatively) clipped fluxes below the 3$\sigma$ level (per channel) before integration. The velocity centroids were determined through a flux-weighted
first moment of the spectrum of each pixel, therefore assigning one velocity to a spectral structure. The dispersion was determined through a
flux-weighted second moment of the spectrum of each pixel. This corresponds to the one-dimensional velocity dispersion (i.e. the FWHM line width of
the spectrum divided by 2.35 for a Gaussian line profile).

\subsubsection{Vibrationally excited HCN}
\label{s:HCN-VIB_mom}

\begin{figure*}
\resizebox{18cm}{!}{\includegraphics[angle=0]{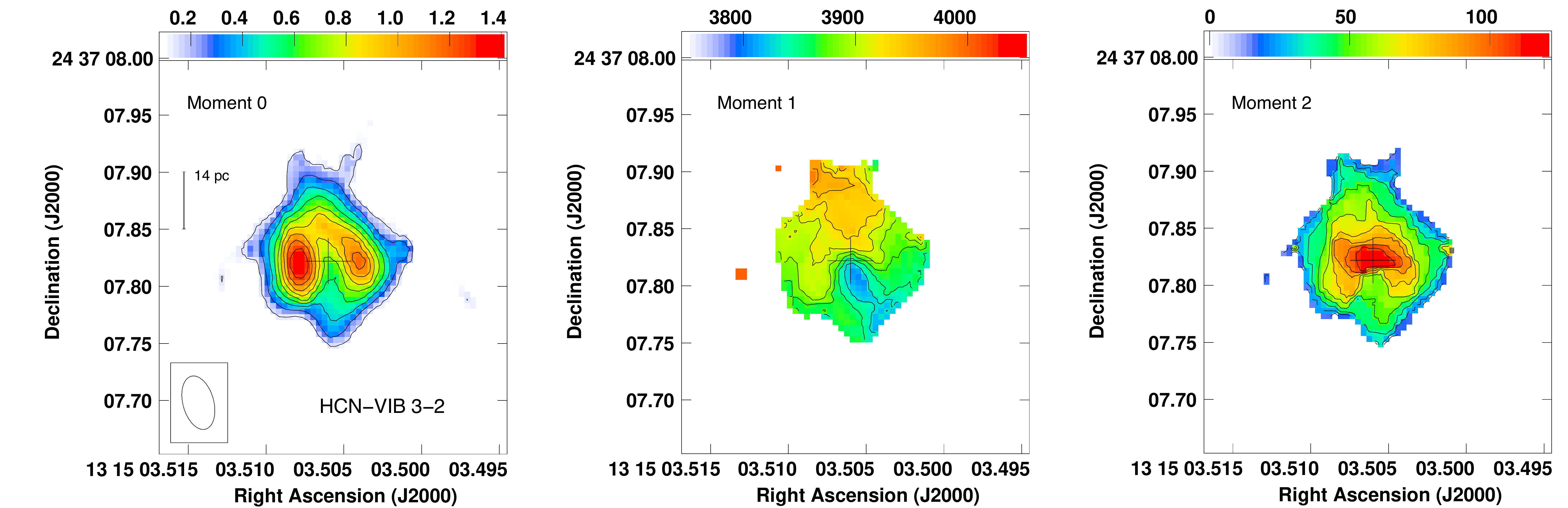}}
\caption{\label{f:vibmomoB6} HCN-VIB 3--2 moment maps. Left: Integrated intensity (mom0) where contours are 0.14$\times$ (1,2,3,4,5,6,7,8,9)  Jy \kms\ beam$^{-1}$. Colours range
from 0.1 to 1.4 Jy \kms\ beam$^{-1}$. Centre: velocity field (mom1) where contours range from 
3830 \kms\ to 3990 \kms\ in steps of 20 \kms\ and colours range from 3750 to 4050 \kms. Right: Dispersion map (mom2) where contours are 
12$\times$(1,2,3,4,5,6,7,8,9) \kms. Colours range from 0 to 120 \kms. The cross marks the position of the 265 GHz continuum peak.}
\end{figure*}

\begin{figure*}
\resizebox{18cm}{!}{\includegraphics[angle=0]{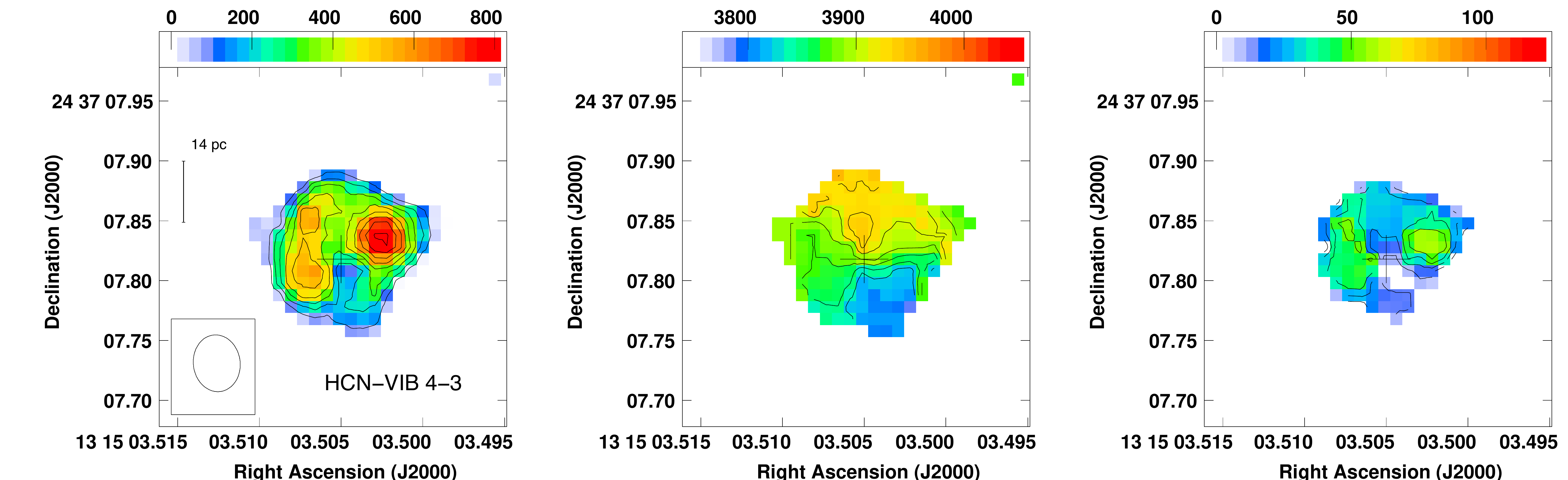}}
\caption{\label{f:vibmomoB7} HCN-VIB 4--3 moment maps. Left: Integrated intensity (mom0) where contours are 0.1$\times$ (1,2,3,4,5,6,7,8,9)  Jy \kms\ beam$^{-1}$. Colours range
from 0 to 1.0 Jy \kms\ beam$^{-1}$. Centre: velocity field (mom1) where contours range from 
3830 \kms\ to 3990 \kms\ in steps of 20 \kms\ and colours range from 3750 to 4050 \kms. Right: Dispersion map (mom2) where contours are 
6$\times$(1,3,5,7,9) \kms. Colours range from 0 to 60 \kms. The cross marks the position of the 350 GHz continuum peak. For better S/N, we present the naturally
weighted HCN-VIB 4--3 moment maps here. }
\end{figure*}

\noindent
{\it Vibrationally excited HCN:} \, We detect luminous HCN $J$=3--2 and 4--3 $\nu_2$=1f (HCN-VIB) emission inside $r$=40 pc of the IC~860 nucleus. The HCN-VIB $J$=3--2 $\nu_2$=1f peak flux density 
above the continuum is $6.5$  mJy\,beam$^{-1}$, which corresponds to a brightness temperature $T_{\rm B}=$115 K. In contrast to the ground-state lines of HCN, the HCN-VIB line is seen
in emission in this region - although on the very nucleus the emission is vanishing for both 3--2 and 4--3 transitions (see below).

The HCN-VIB moment 0, moment 1, and moment 2 maps are presented in Figs.~\ref{f:vibmomoB6} and~\ref{f:vibmomoB7}. For better S/N, the HCN-VIB 4--3 moment
maps are shown in natural weighting (with beam size $0.\asec 047 \times 0.\asec 039$).
All three moment maps cover the same velocity range: for HCN-VIB 3--2 we integrated from 3650 to  4100 \kms\ while for the HCN-VIB 4--3 moment
maps we had to cut out velocities greater than 3990 \kms\ due to a blend with the vibrationally excited HC$_3$N $J$=39--38, $\nu_7$=1 line. 
It has an expected peak intensity of about 50\% of that of the HCN-VIB line and appears as a red-shifted shoulder
to the HCN-VIB 4--3 line. 
The HCN-VIB 3--2 line is not affected by blending with HC$_3$N. There is a small emission feature
(less than 1/5th of the HCN-VIB peak at $\Delta V$=+390 \kms\ from the HCN-VIB peak) 
that could be excited CH$_3$OH or vibrationally excited H$^{13}$CN $\nu_2$=1. 

\smallskip
\noindent
{\it Vibrationally excited HC$_3$N:} \, Vibrationally excited HC$_3$N $J$=29--28, $\nu_7$=2 line emission is located +479 \kms\
from the line centre of the ground state HCN 3--2 line. We also detect luminous, vibrationally excited HC$_3$N J=12-11, $\nu_7$=1 and $\nu_6$=1
emission at $\lambda$=3~mm. We defer the discussion of the vibrationally excited HC$_3$N emission to a later paper. 

\smallskip
\noindent
{\it Moment 0 maps:} \, The 3--2 and 4--3 HCN-VIB structures share a similar double-peaked structure. For the 3--2 line the integrated emission peaks are found along an east--west axis, perpendicular to
the nuclear rotation direction. For the 4--3 line the brightest is 0.\asec 04 (11 pc) to the northwest (position angle (PA) of -45$^{\circ}$) of the continuum peak and a weaker one is 
0.\asec 04 southeast  (PA=135$^{\circ}$).  However, the band~7 HCN-VIB line may be affected by the blend with vibrationally excited HC$_3$N. The distance between the 3--2 line peak and the centre
is shorter - 0.\asec 025. For both lines the emission is strongly depressed (almost down to zero) on the nucleus. Also for both lines the emission is fainter to the south than to the north and the source size
(from a two-dimensional Gaussian fit to the intensity distributions) is 0.\asec060  for both transitions. Since the intensity structure is not single peaked, such a fit provides only an estimate of the source-size of the emission.

\smallskip
\noindent
{\it Moment 1 maps:} \, Both the 3--2 and 4--3 lines show a clear north--south velocity gradient in the inner 0.\asec070  (14 pc) with projected maximum, intensity weighted velocities of $\pm$100 \kms. The 3--2 velocity
structure  also shows deviations from circular rotation along the minor axis, and a shift of position angle from PA=25$^{\circ}$ at $r$=0.\asec 1 to PA=0$^{\circ}$ at $r$=0.\asec 05 to 0.\asec 1 .

\smallskip
\noindent
{\it Moment 2 maps:} \,  The 4--3 dispersion map has two distinct maxima that overlap in position with the moment 0 peaks, although both extend more to the north than the integrated intensity peaks.  For the 3--2 map,
the high dispersion region extends along the minor axis with a peak in the centre. The  zero-intensity full line width $\Delta v$ peaks at 250 \kms, (for the 3--2 line). The dispersion maximum occurs in the northwestern
emission peak for the 4--3 line,  and in the centre and eastern intensity maximum for the 3--2 line. In general, the dispersion peaks are roughly perpendicular to the rotation seen in the moment 1 map.


\subsubsection{Ground-state lines}
\label{s:ground}

\begin{figure*}
\resizebox{18cm}{!}{\includegraphics[angle=0]{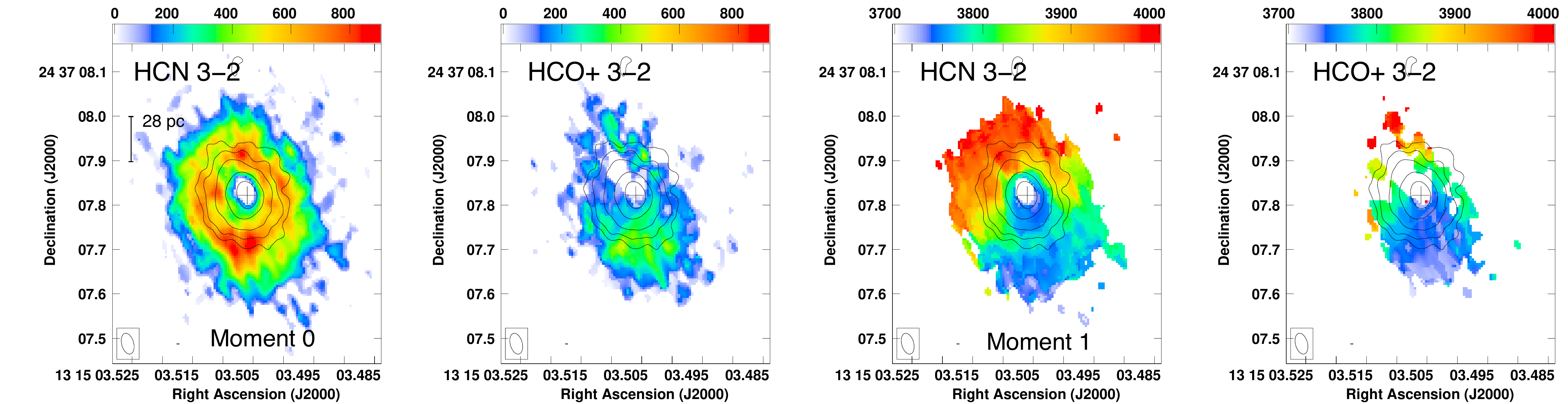}}
\caption{\label{f:groundmom0}  Band~6 integrated intensity (moment 0) and velocity field (moment 1) maps (of emission only) of the ground-state HCN and HCO$^+$ 3--2 lines. 
{ We note that the  white `hole' in the centre is due to absorption and that absorption affects the line profiles out to at least 50\% of the continuum. Some of the `missing' emission in structures 
is also due to line self-absorption.}. The HCN and HCO$^+$ 3--2 line emission is overlayed on the 1~mm
continuum contours ($0.5 \times (1,2,4,8,16)$ mJy). The cross marks the position of the 265 GHz continuum peak. Left panel: HCN 3--2 moment 0 map where colours range 
from 0 to 0.9 Jy \kms\ beam$^{-1}$; Left centre panel: HCO$^+$ 3--2 moment 0 map with colours  ranging from 0 to 0.48 Jy \kms\ beam$^{-1}$. Right centre panel: HCN 3--2 moment 1 map. Right panel:
HCO$^+$ 3--2 moment 1 map. Colours range from 3700 to 4000 \kms.}
\end{figure*}

\begin{figure}
\resizebox{9cm}{!}{\includegraphics[angle=0]{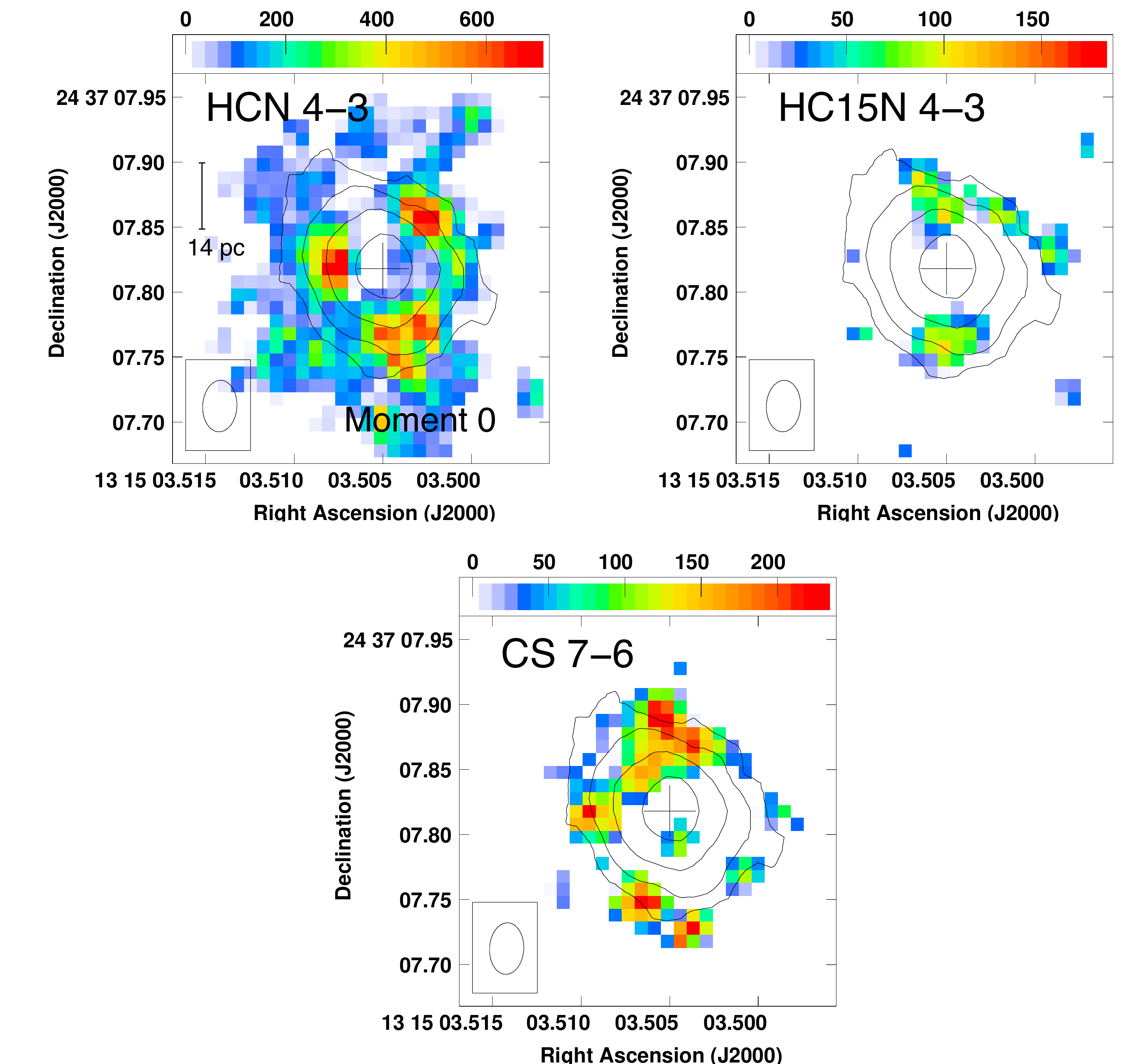}}
\caption{\label{f:groundmom_B7}  Band 7 moment 0  maps (of emission only) of three ground-state lines overlayed on the 343 GHz continuum contours (0.74$\times$(1,2,4,8) mJy). We note that the  white `hole' in the centre is due to absorption and that absorption affects the line profiles out to at least 50\% of the continuum. Some of the `missing' emission in structures 
is also due to line self-absorption. The cross marks the position of the
continuum peak. Left panel: HCN 4--3 where colours range from 0 to 0.7 Jy \kms\ beam$^{-1}$; Centre panel: HC$^{15}$N 4--3 with colours  ranging from 0 to 0.18 Jy \kms\ beam$^{-1}$; Right panel: CS 7--6 with
colours ranging from 0 to 0.23 Jy \kms\ beam$^{-1}$.}

\end{figure}

We present moment 0 and moment 1 maps (Figs.~\ref{f:groundmom0} and~\ref{f:groundmom_B7}) for selected ground-state lines detected in bands 6 and 7. All moment maps cover
the same velocity range: 3650 to  4100 \kms. The lower-energy (normally collisionally excited) ground-state lines are seen
in a mixture of emission and absorption; we note that the moment maps presented here only cover the emission. Therefore  absorption regions will appear white since the absorption depth and structure is not shown.

\smallskip
\noindent
{\it Band 6 and 7 (1 and 0.8 mm):} \, We detect strong emission/absorption from ground-state lines of HCN,  HC$^{15}$N, HCO$^+$, 3--2 and 4--3, CS  7--6 and CH$_2$NH $4_{1,3} - 3_{1,2}$. The average
spectrum of the inner 50 pc reveals a reversed P-Cygni profile as previously reported for HCN and HCO$^+$ 3--2 in \citet{aalto15b}.  However on smaller scales the emission/absorption behaviour changes, and is
different for different molecules and transitions. On the mm continuum peak (see Fig.~\ref{f:all_spectra}) all lines are seen in absorption: HCO$^+$ and HCN 3--2  show a shift (relative to systemic velocity $v_{\rm sys}$) 
of  +50 \kms\ with a red wing;  HCO$^+$ and HCN 4--3 show little 
shift from $v_{\rm sys}$ (we note that the HC$_3$N 39--38 line contaminates HCN 4--3 on the blue side (-155 \kms)); 
CS 7--6 is blue-shifted from $v_{\rm sys}$ with a blue wing extending out to -200 \kms\ and HC$^{15}$N also shows a small blueshifted absorption peak on the continuum peak similar to CS 7--6. 
The interpretation of the  absorption line shapes is discussed further in Sect.~\ref{s:inflow}. 
We note that both ground-state HCN lines should be contaminated by  a strong, $\nu_2=1e$ vibrationally excited HCN emission close to their line centres. The line should be as strong as the $\nu_2=1f$ line, but
is not visible in the absorption spectra of HCN 3--2 or 4--3. This suggests that the photons from the $\nu_2=1e$ line are being absorbed by foreground HCN.

\begin{enumerate}
\item{\bf HCN and HC$^{15}$N moment maps: } \, HCN 3--2 emission is found in the inner 0.\asec 5 (140 pc)  in a somewhat peculiar `box-like' distribution. Emission is suppressed and distorted by absorption
in the inner 0.\asec 2 - 0.\asec 3 and completely absorbed in the very inner 0.\asec070. The velocity field outside the distorted region suggests rotation with PA =20$^{\circ}$-35$^{\circ}$. The full absorption extends
to the north with a PA=25$^{\circ}$. 

Patchy HCN 4--3 emission is found in the inner 0.\asec 25. Emission is suppressed in a region to the northeast of the nucleus (similar to HCN 3--2).  HC$^{15}$N 4--3 emission is poorly correlated to that of HCN 4--3.
This is to be expected if there are strong effects of opacity and absorption.  Red- and blueshifted HC$^{15}$N emission is found 0.\asec 05 to the north and south of 
the 0.8~mm continuum peak. The velocity shift is 250 \kms\ and is consistent with velocities found for the HCN-VIB north--south rotation. \\

\item {\bf HCO$^+$ and CS: } \,  HCO$^+$ 3--2 emission is located inside 0.\asec 5 (140 pc)  in a  distribution that is dominated by emission to the south. Emission is suppressed and distorted by absorption in the inner 0.\asec 2 - 0. \asec 3
and completely absorbed in the inner 0.\asec050 , that is the northern part of the inner peak.  There is also absorption in a narrow structure extending 0.\asec 1 (29 pc) to the north with PA=25$^{\circ}$.  The velocity field outside the distorted
regions suggests rotation with a similar PA to that of HCN 3--2.  No HCO$^+$ 4--3 emission was found above the noise and therefore no moment 0 map was made. 

CS 7--6 emission is found in the inner 0.\asec 2 and it lacks the suppression to the north seen for HCN and HCO$^+$ 4--3 and 3--2. The intensities are  in general about one third of those found for HCN 4--3.

\end{enumerate}

\smallskip
\noindent
{\it Band 3 (3 mm): } We detect line emission from CS 2--1, $^{13}$CO 1--0, C$^{18}$O 1--0, HC$_3$N 12--11, 11-10 HNCO, and CH$_3$CN. The line shapes are strongly affected by continuum- and self-absorption (apart from the
vibrational lines).  We defer the discussion of the interesting band 3 line emission to an upcoming paper, and focus on discussing the band 3 continuum in this paper.


\section{Discussion}


\subsection{An opaque, dusty nucleus}

Combining our observations at Q~band (45 GHz), ALMA band~3 (100 GHz), and ALMA band~6 (260 GHz) leads us to the conclusion that the mm-wavelength continuum emission is dominated
by dust. Below we discuss the potential contribution from other emission mechanisms.

\subsubsection{Contribution from synchrotron and free-free emission}
\label{s:synchrotron}

{\it Synchrotron emission} \, At longer wavelengths (L and C band - 1.6 and 5 GHz) the continuum of IC~860 has an almost flat spectrum with spectral index $\alpha=-0.33$ \citep{baan06} which may raise
concerns that synchrotron emission may contaminate emission at mm wavelengths. However, our VLA Q-band data point shows that the spectrum steepens towards shorter wavelengths, and that
contamination will be very small. This is consistent with our (lower resolution) C (5 and 6 GHz) and K (18, 19 and 20 GHz) VLA observations where we find a spectral index between C and K of $\alpha=-0.6$. 
These observations will be presented in a forthcoming paper. In addition, the high-resolution C-band continuum image by \citet{baan17} reveals that the structure  is extended on scales of 0.\asec 5
 - larger than the mm-continuum sizes we find here.  Only a small fraction of the C-band  flux (4-5 mJy) can be found on scales of the mm-wave dust peak.  This further supports the notion that synchrotron
 contamination to the mm dust continuum is minimal.

{\it Free-free emission} \, Assuming that the Q~band core flux consists entirely of  optically thin free-free emission, we can project its expected flux contribution to higher frequencies:  a maximum of 
0.46 mJy at 3~mm and 0.42 at 1~mm, i.e. less than 14\% at band~3 and 3\% at band~6. 

There is however a possibility that  optically thick  free-free emission from an ionized region of very dense ($n>10^6$  $\cmmd$) gas may be hiding at the core of IC~860. It is not inconceivable that such
a source (with brightness temperature $T_{\rm B}=5 \times 10^3 - 10^4$ K) might exist near an AGN or a (very) compact starburst.  We use the high-resolution band~C observations of \citet{baan17} to
estimate the core spectral index, between C and Q-band, to $\alpha=-1.2$. This is close to the maximum (in absolute value) observed in galaxy nuclei \citep{odea98} and consistent with synchrotron losses
resulting in the steepening of the spectrum at frequencies $\nu > 1$ GHz. Future high-resolution multi-frequency studies around 30-50 GHz will provide tighter constraints on the low-frequency SED and the
balance between synchrotron, free-free, and dust emission. Based on current information, we conclude that there is no remaining free-free Q-band flux (optically thick or thin) that can lead to significant
contributions at mm wavelengths.

\subsubsection{Dust temperature and opacities}
\label{s:dust_opacity}
Provided that the 3 and 1~mm continuum emission mostly stems from dust, the opacities at mm wavelengths are likely significant. For the 1~mm continuum brightness of $T_{\rm B}\sim$250 K, the dust
temperature would be unrealistically high ($\gapprox 2000$ K) for optically thin emission (see e.g. \citet{sakamoto17}). The opacity must be unity ($T_{\rm d}$=400 K) or higher, and the $N$(H + H$_2$) associated with
optically thick mm dust continuum is high.

According to the formalism of \citet{keene82} and \citet{hildebrand83},  the relation between opacity and column density is $N$(H + H$_2$)/$\tau_{\lambda}$=$1.2 \times 10^{25}(\lambda/400 \mu{\rm m})^2$ $\cmmt$,
and for  $\tau$=1 at $\lambda$=1~mm we find a column density of  $N$(H + H$_2$)$\sim10^{26}$ $\cmmt$.   Using a modified black body to produce a SED\footnote{We used a modified black body
to produce the SED. We calculate an optical depth at each wavelength and then we determine the dust temperature (modified by the optical depth). The mass absorption coefficients used can be found in \citet{gonzalez14}. 
For a standard dust-to-gas ratio of 1/100,  the H$_2$ column densities are $N$(H$_2$)=$5 \times 10^{26}$ $\cmmt$. This implies an average gas number density of $n \approx 10^7$ $\cmmd$.}, we fit a  
$T_{\rm d} \simeq$280 K and $\tau$ of $\sim$5 (at 1~mm) and an extreme column density of $N$(H + H$_2)\simeq5 \times 10^{26}$  $\cmmt$ (assuming a standard dust-to-gas ratio of 1/100). 
This implies an average gas number density of $n \approx 10^7$ $\cmmd$. 

The high $T_{\rm d}$ of the mm-emitting dust core means that it should emit strongly in the MIR. Our model over-predicts the observed  14~$\mu$  MIR flux \citep{lahuis07} (90~mJy) by a factor of $\sim$100-200,
corresponding to a $\tau \sim 5$ in the MIR. This implies that a column density of at least $N$(H$_2$) $\sim10^{23}$ $\cmmt$ is foreground to the hot, 280~K dust core. 
\citet{lahuis07} detect MIR HCN 14~$\mu$m absorption with $T_{\rm ex}\sim$280 K.  This absorption may be occurring in gas in front of the opaque mm-core and/or in hot gas associated with other MIR
structures of lower opacity such as in an outflow or in a star forming region. Our model assumes a simple, smooth non-clumpy  structure of a single temperature, but the actual situation is likely more complex, requiring a more sophisticated approach in the future. 

The model luminosity for a spherical distribution with the fitted $T_{\rm d}$ is $4.3 \times 10^{11}$ \lsun\ and $1 \times 10^{11}$ \lsun\ for a thin disk of the same dimensions. However, due to the high opacities,
continuum photons may become trapped, elevating the internal $T_{\rm d}$ and increasing the volume of hot gas \citep[e.g.][]{kaufman98, rolffs11b}. Hence, to determine the true luminosity of the inner structure, the
trapping effects must be taken into account (e.g. Gonzalez-Alfonso et al. in prep.). We discuss this further in Sect.~\ref{s:buried}.

\subsubsection{Are the dust properties unusual?}
The inferred column density of $N$(H + H$_2$)=$5 \times 10^{26}$ $\cmmt$ is very high and renders the IC~860 nucleus extremely opaque, strongly suppressing X-ray, IR, and even mm emission. Similarly
large values for $N$(H + H$_2$) have been suggested for the ULIRG Arp~220  \citep{scoville17}. However, are there alternative possibilities to the vast H$_2$ columns?

\begin{enumerate}

\item{\it Elevated dust-to-gas ratio:} \, There is some evidence of high dust-to-gas ratios in some ULIRGs or in some dust-reddened quasars \citep{wilson08, banerji17}. Values of 1/30 have been found in
the lenticular galaxy NGC5485 \citep{Baes14}, although this dust does not appear to be associated with molecular gas.   IC~860 is a post-starburst galaxy, so it could be speculated that the starburst resulted in
a higher dust-fraction in the gas which is now inflowing to the central region of IC~860. However, studies of the yields from evolved AGB stars imply canonical dust-to-gas ratios \citep{dharmawardena18}. 
An issue is also how the dust-to-metal ratio is evolving with time and affected by starbursts (see e.g. a recent study by \citet{devis19}). However, in dense regions grain growth may occur on short timescale
 \citep{debennassuti14} that can impact dust-to-gas ratios.\\

\item{\it Unusual dust-grain properties:}  \, Very large, mm-sized dust grains may also provide high opacity at mm wavelengths without an associated, extreme $N$(H$_2$). Studies of MIR to mm-wave dust continuum
in protoplanetary disks seems to allow for the possibility of grain-growth at the outskirts of the disks \citep[e.g.][]{lommen10,l_ricci12}. However, these processes occur on small scales in cold, planet-forming structures around stars.

\end{enumerate}

We conclude that a higher-than-normal dust-to-gas ratio in the centre of IC~860 is possible and requires further study. There is however currently no direct evidence to support this, and we will therefore continue to adopt a
standard dust-to-gas ratio of 1/100. The high temperatures of the dust make significant ensembles of mm-sized grains unlikely. 


\subsection{The HCN-VIB emission - structure and dynamics of the hot nuclear gas}
\label{s:hcn-vib}

Intense HCN-VIB emission is found in the inner 0.\asec 1, where the ground-state HCN lines are almost entirely seen in absorption (see e.g. Fig.~\ref{f:pV}).  HCN-VIB line-to-continuum ratios peak 30 mas (9 pc) from the centre along 
the major axis (and exceed 100 K). Closer to the centre, the line-to-continuum ratios start to drop. In Sect.~\ref{s:dust_opacity} we discuss the origin of the mm continuum and conclude that it is  emerging from hot
($\simeq$280 K) opaque dust (but also suggest that the possibility of the existence of a nuclear, very dense plasma should be investigated in the future).  The suppression of the HCN-VIB emission towards the continuum
peak is expected in this scenario. If the dust opacity is $\gapprox$1, then only very little emission will emerge from the core. 
In addition, if the excitation temperature of the HCN-VIB emission matches that of the continuum, then the line will vanish in front of the continuum peak. If we compare the HCN-VIB brightness temperature, 
$T_{\rm B}$(HCN-VIB), to the continuum  temperature, $T_{\rm B}$(continuum), we find that this approach works well for the inner 0.\asec 1 of IC~860; it implies an excitation temperature  $T_{\rm ex}$(HCN-VIB)
of $\sim$200 K - which is not unexpected given the high value of $T_{\rm d}\simeq$280 K. The gas densities in this region are estimated to $n=10^7$ $\cmmd$ for a normal dust-to-gas ratio (see Sect.~\ref{s:dust_opacity}) which
means that gas and dust will be thermalized to the same temperature. Cooler foreground gas will be seen in absorption, while hot, dense gas near the dusty nucleus (or within it) risks having its brightness temperature
severely reduced (or vanished). The opacity in the HCN-VIB line must be significant since $T_{\rm B}$(HCN-VIB) is high,  further supporting the notion of large gas column
densities in the core of IC~860.  The fact that $T_{\rm B}$(HCN-VIB) peaks perpendicular to the north--south major axis may be a result of higher opacities along the major axis.

\subsubsection{HCN-VIB kinematics and nuclear dynamical mass}
\label{s:hcn-vib_dyn}

\begin{figure}
\resizebox{9cm}{!}{\includegraphics[angle=0]{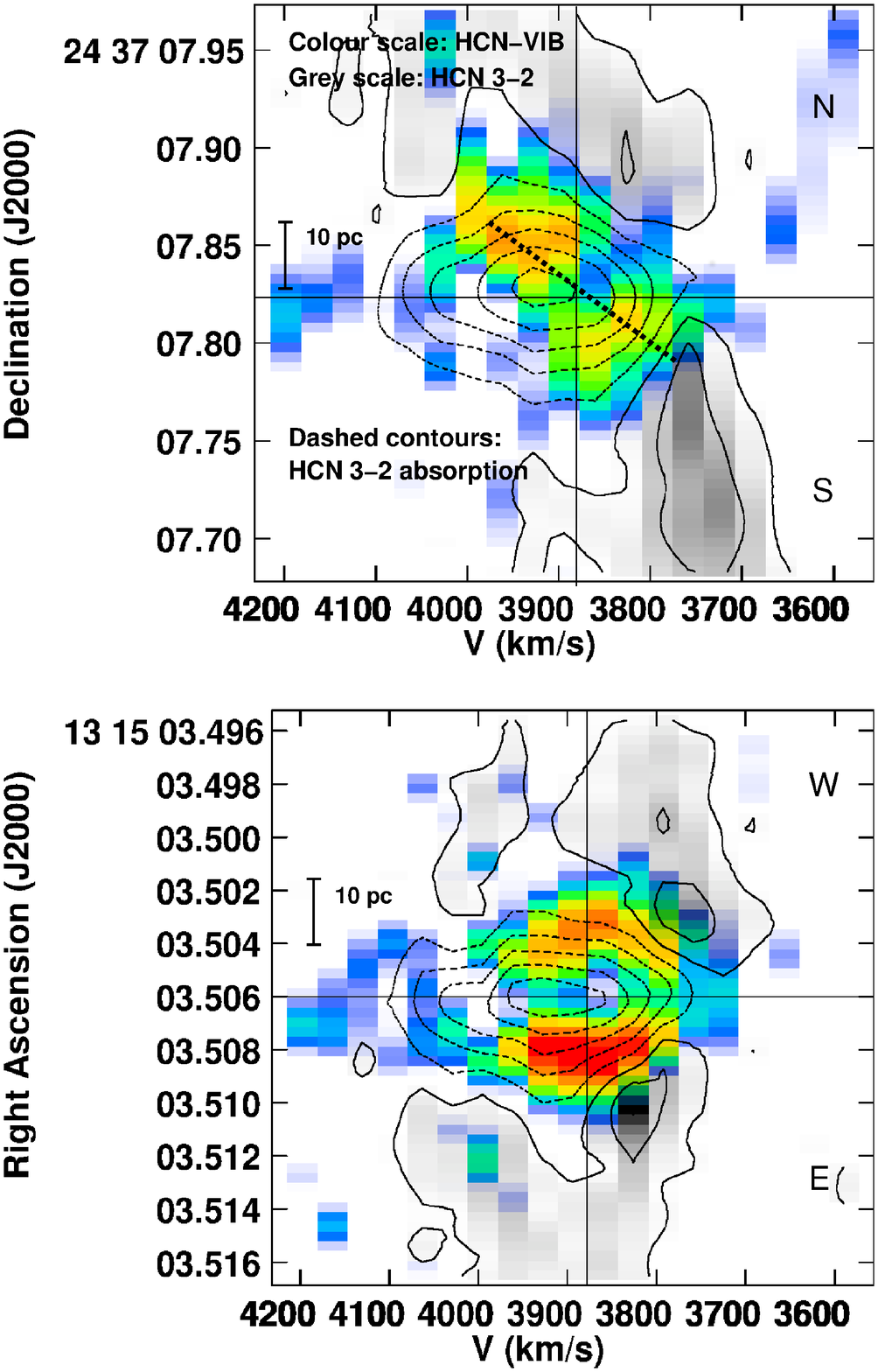}}
\caption{\label{f:pV}  Position velocity (pV) diagrams of HCN-VIB and HCN 3--2.  Colours indicate HCN-VIB and greyscale indicates HCN 3--2 emission - both range from 1 to 6.5 mJy beam$^{-1}$. The
contour range is 1.2$\times$(-9,-7,-5,-3,-1,1,3). HCN 3--2 absorption is indicated by negative, dashed contour lines. Top panel: Cut is along the north--south (PA=0$^{\circ}$) major axis of the nuclear disk. The black dashed line
indicates the intensity weighted rotational velocity within $r$=10 pc. Lower panel: Cut is along the east--west (PA=90$^{\circ}$) minor axis of the nuclear disk. The  $v_{\rm sys}$  of 3880 \kms\ is indicated with a vertical black line.
The position of the band 6 continuum is indicated by a horizontal black line.
}
\end{figure}

The nuclear, rotational major axis is along PA=0$^{\circ}$ and in the moment maps the peak rotational velocity is 100 \kms\ (Sect.~\ref{s:HCN-VIB_mom}). Figure~\ref{f:pV} shows the position--velocity (pV) diagram along the nuclear major axis. The figure includes the HCN-VIB emission as well as the ground-state HCN 3--2 emission and its nuclear absorption. 

The pV diagram shows that the HCN-VIB line widths are broad with $\Delta$v around 150 \kms.  Non-circular blueshifted emission can be seen to the  northwest in the pV diagram. Intensity-weighted projected rotational velocities at $r$=10 pc are 80 \kms\ to the north and 100 \kms\ to the south.  However, due to the high line widths and the potential $T_{\rm ex}$ suppression, we should view these numbers with caution. We see that the systemic velocity, $v_{\rm sys}$, 
of 3880 \kms\ (marked in the figure) from \citet{aalto15b} fits reasonably well with the HCN-VIB dynamics.  

The pV diagram shows that the HCN-VIB appears in emission in the inner region, where the HCN 3--2 line is dominated by absorption. The HCN-VIB line is found in emission all the way into the centre, but its line-to-continuum
ratio is reduced (see Sect.\ref{s:hcn-vib}) in the central 20 mas. Background blueshifted HCN 3--2 emission can be seen to the south. To the north, HCN-3--2 emission lines are broad and emission from non-circular motions can also
be seen on the blue side. The absorption peak is shifted to the red by 40-50 \kms\ from $v_{\rm sys}$, a shift that is also seen on  larger scales (Sect.~\ref{s:ground}).

\smallskip
The 0.8-3~mm continuum is almost circular despite the large nuclear line-of-sight $N$(H$_2$) column density. The major/minor axis ratio implies an inclination of $i$=30$^{\circ}$ - 40$^{\circ}$, but a more inclined disk combined with
a minor axis outflow will also result in a structure that appears more face-on than it actually is  (e.g. in Arp~220; \citet{sakamoto17, barcos18}). If we assume that the structure is a near face-on disk with $i$=30$^{\circ}$,  and a projected
rotational velocity of 100 \kms, then the enclosed mass inside $r$=10 pc can be estimated to  $M_{\rm dyn}=9 \times 10^7$  \msun\footnote{A simple estimate of the dynamical mass is  $M_{\rm dyn}=2.3 \times 10^8 \times (v_{\rm rot}/100)^2 \times (r/100)$ \msun. Here $v_{\rm rot}$ is the rotation speed in \kms\ and $r$ is in pc.}. For a disk inclination of $i>60^{\circ}$ (the estimated large-scale inclination of IC~860),  the dynamical mass of the $r$=10 pc disk is  instead $M_{\rm dyn} \lapprox 3 \times 10^7$  \msun. In the
discussion on the nature of the buried activity in Sect.~\ref{s:buried}, we adopt an upper limit on the dynamical mass of $M_{\rm dyn} \lapprox 9 \times 10^7$  \msun\ - resulting in lower limits on the luminosity-to-mass ($L/M$) ratios. 

In the following section (Sect.~\ref{s:inflow}) we discuss the layered  red- and blueshifted absorption of the ground-state lines and their implication for a "near face-on" ($i$=30$^{\circ}$) and "near edge-on" ($i \gapprox60^{\circ}$) orientation of the dusty nucleus of IC~860.


\subsection{Foreground gas: inflow, outflow, and the structure of the IC~860 dusty nucleus}
\label{s:inflow}

In Sect.~\ref{s:ground} we present the line shapes of the ground-state HCN, HCO$^+$, and CS lines (Fig.~\ref{f:all_spectra}) averaged in the inner 0.\asec 1. The
HCN and HCO$^+$ lines show average, redshifted,  reversed P-Cygni profiles. These redshifted absorption profiles are also seen on small scales in front of the nucleus (Fig.~\ref{f:pV}).
In contrast , CS 7--6 has a blueshifted line profile.  

\subsubsection{Evidence of foreground inflowing gas}

In the moment maps  (Figs.~\ref{f:groundmom0} and ~\ref{f:groundmom_B7})  we see that the HCN and HCO$^+$ lines are affected by absorption:  extending from the mm continuum
peak 0.\asec 2 (57 pc) to the northeast. For HCO$^+$ 3--2, this absorption is a striking, narrow (15-20 pc across) structure where the emission completely vanishes, while the  HCN 3--2 
emission is suppressed but does not completely vanish. 
Foreground  ground-state HCN and HCO$^+$ absorb the intense continuum emission in the centre, and towards the northeast they may also suffer from line-of-sight absorption by foreground gas which 
removes emission in the northeast. This is supported by  HC$^{15}$N 4--3 being seen in emission where HCN 4--3 is absent (Fig.~\ref{f:groundmom_B7}). 

The ground-state HCN and HCO$^+$ lines show redshifted, reversed P-Cygni line profiles in the inner 0.\asec 1 (Fig.~\ref{f:all_spectra}) and also on smaller scales in front of the nucleus 
(Fig.~\ref{f:pV}). Such line shapes are often taken as evidence of infall or inflow, but they may also be caused by foreground tidal structures above the galaxy, or gas on larger-scale non-circular orbits.
However, here we find that the foreground absorption appears to end in the nucleus, which suggests that they are not chance crossings of gas above the plane of the galaxy. In addition, the
foreground absorption is roughly aligned with the larger-scale stellar bar which is likely responsible for funneling gas to the centre. The lack of a corresponding 0.\asec 2 HCN/HCO$^+$ absorption
structure to the southwest may be because the gas is flowing from both the northeast and the southwest where the northeast flow is foreground to the core and the southwest gas is in the background. Evidence of this
can be seen in the pV diagram (Fig.~\ref{f:pV}) where blueshifted
HCN 3--2 emission can be seen behind the continuum source (albeit not in the very inner 50 mas which may be due to high opacities in the continuum source).

\subsubsection{A dense outflow?}
 
\begin{figure}
\resizebox{7cm}{!}{\includegraphics[angle=0]{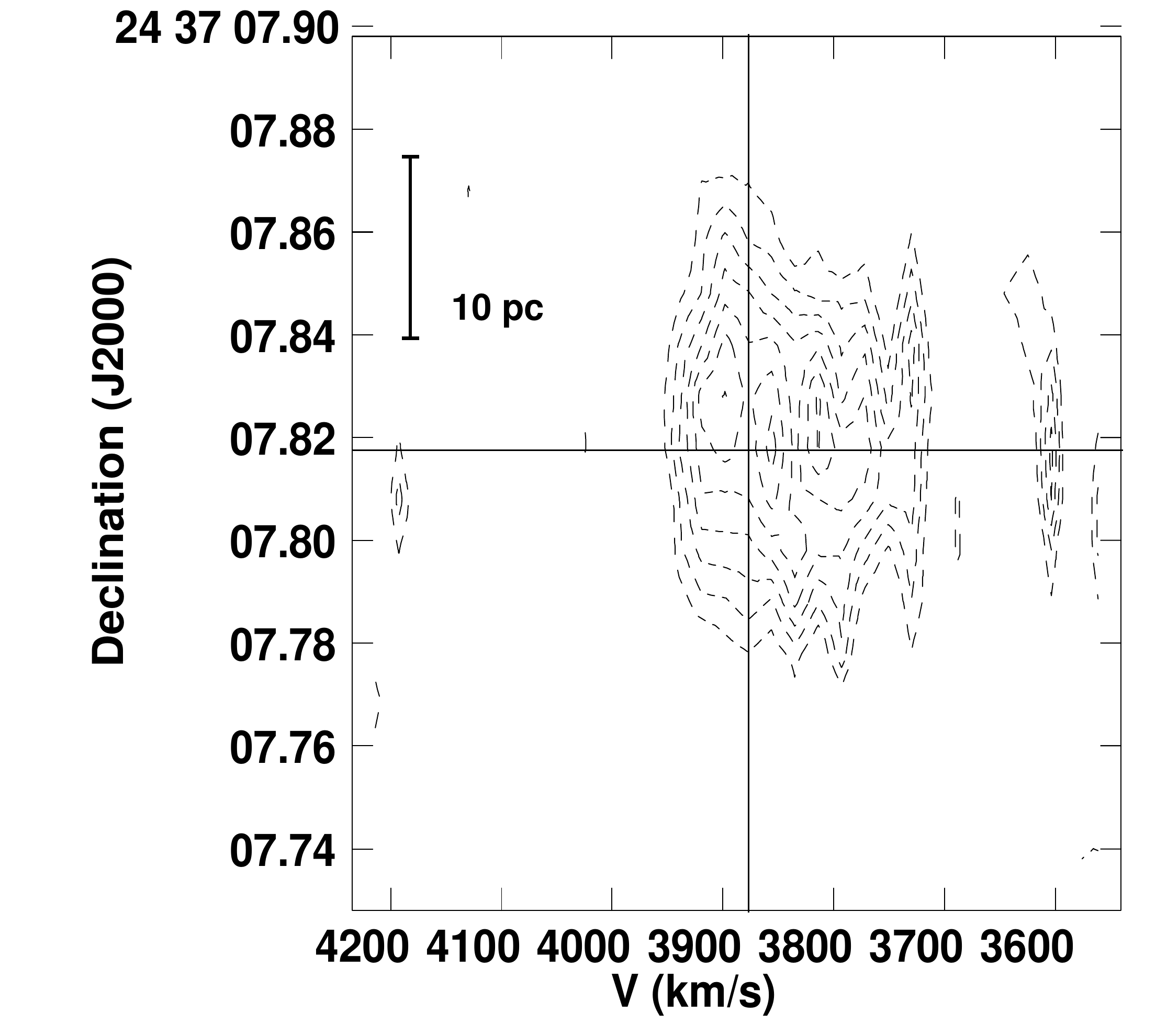}}
\caption{\label{f:CS_pV}   Position velocity (pV) diagram of CS 7--6 absorption, indicated by negative, dashed contour lines. The contours are 0.8 $\times$(-8,-7,-6, -5,-4,-3,-2) mJy and the
cut is along the north--south (PA=0$^{\circ}$) major axis of the nuclear disk. The  $v_{\rm sys}$  of 3880 \kms\ is indicated with a vertical black line.
The position of the band 7 continuum is indicated by a horizontal black line.
}
\end{figure}

 In contrast to HCN and HCO$^+$ 4--3 and 3--2, CS 7--6 has a blueshifted absorption profile (Fig.~\ref{f:all_spectra}).  Just like HCN/HCO$^+$ it is completely absorbed towards the continuum
 peak in the centre, but does not show the same extended absorption pattern towards the northeast. The same appears to be the case for HC$^{15}$N 4--3.  In Fig.~\ref{f:CS_pV} we show the 
 high-resolution CS 7--6 north--south pV diagram, which reveals centrally concentrated, blueshifted absorption out to velocities of 3700 \kms\ (-180 \kms\ from $v_{\rm sys}$) with a possible additional
 feature at 3600 \kms.
 
 The blueshifted CS 7--6 profile may indicate a compact outflow, or the base or launch region of a larger-scale outflow. The critical density of the CS 7--6 transition matches the inferred average density
 of the inner structure suggesting a very dense outflow with $n>10^6$ $\cmmd$; however, the excitation may be dominated by radiation rather than collisions here.
 The CS 7--6 line is more highly excited ($T_{\rm L}\simeq$49 K) than HCN and HCO$^+$ 4--3 ($T_{\rm L}\simeq$25 K), 
 and is therefore expected to be somewhat more centrally concentrated (albeit this should be a small effect). The absorption depth in front of the continuum is lower for CS 7--6 compared to HCN and
 HCO$^+$ 4--3, The blueshifted northwestern HCN 3-2 emission,  from gas on non-circular orbits,  may be an extended part of this outflow, or other non-circular motions in this
 complex nucleus. 

\subsubsection{The structure of the dusty core of IC~860}

\begin{figure}
\resizebox{9cm}{!}{\includegraphics[angle=0]{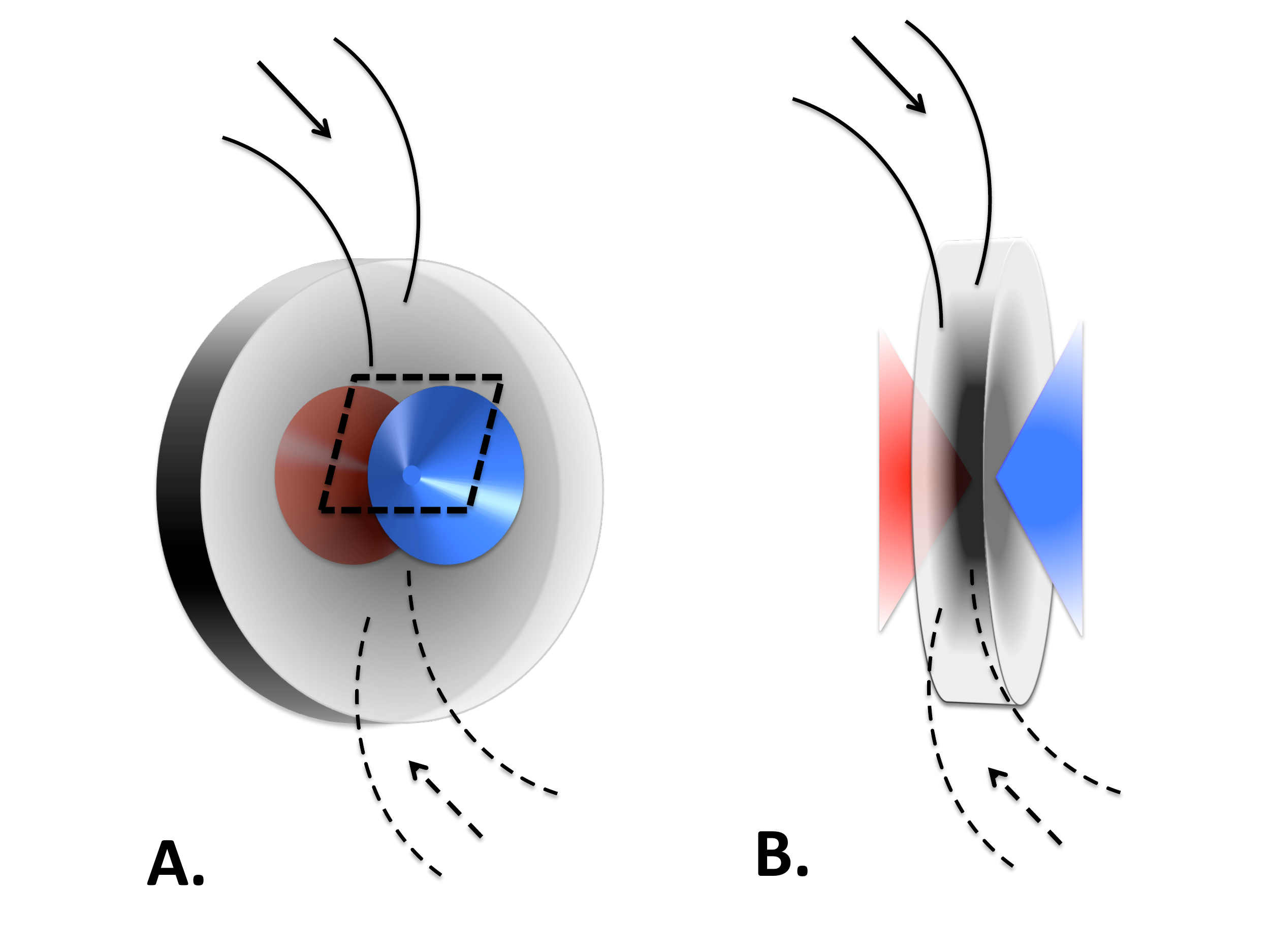}}
\caption{\label{f:cartoon}  Cartoons of two scenarios of the nuclear structure: \textbf{A} "Near face-on scenario" with the inflow structures marked with curves and arrows, and a nearly head-on blueshifted outflow structure, and
a redshifted flow directed away from us, likely obscured by the disk. The dashed
parallelogram marks foreground material that obscures the outflow in HCN and HCO$^+$. \textbf{B} "Near edge-on scenario" where the inflowing gas connects to a heavily inclined disk.  The blue-shifted outflow may partially be
obscured by HCN and HCO$^+$ gas flows in the outskirts of the inclined disk, but not fully. Additional obscuring material is likely necessary.  For both figures, east is to the left and north is up.}
\end{figure}

Combining the emission and absorption line structures suggest a central morphology of a relatively large-scale ($>$50 pc) inflow structure that feeds the central dusty core. 
In Fig.~\ref{f:cartoon} we present simple cartoons where scenario {\bf A} is a near face-on structure and scenario {\bf B } is near edge-on. Both require an inflow component - either directly connecting
with the nucleus, or via an inclined disk - as well as a dense, compact outflow.

\begin{description}

\item{\it Scenario A - slightly inclined disk:} \,  If we assume that the structure is a near face-on disk with $i$=30$^{\circ}$,  we have to add an additional component of foreground gas
 along PA=25$^{\circ}$ from the north and into the nucleus  to explain the lanes of absorbed ground-state lines of for example HCN and HCO$^+$.  A foreground screen of colder dust is also required to absorb
 most of the MIR emission from the hot $T_{\rm d}\simeq230$ K nuclear dust. The blueshifted outflowing gas is oriented close to the line-of sight towards us and is hidden behind layers of foreground gas and/or dust. 

\smallskip 
\item{\it Scenario B - near edge-on disk:} \, In this scenario, the foreground structure along PA$^{\circ}$=20-25$^{\circ}$  connects to an inclined ($i\gapprox60^{\circ}$) disk, within which the gas may continue to
reach the very nucleus. The large column density in the inclined disk will contribute to suppressing HCN-VIB emission along the major axis. An inclined disk can also produce the inferred large $N$(H$_2$) towards
the nucleus without an additional structure. Gas in an outflow will be exposed to polar MIR emission that can excite the minor-axis HCN-VIB emission.
The velocity dispersion will peak along the minor axis - even if the line core velocity shifts are small (due to the outflow axis being near edge-on - but also due to moderate outflow velocities).

\end{description}

\noindent
Both  scenarios have caveats, and further studies will be required to establish the best model for the nucleus of IC~860. However, scenario B does not require additional foreground and background components and is
the simplest model that can explain our observations.

\subsubsection{Inflow and outflow velocities}
\label{s:inflow_rate}

Determining the inflow velocity in the larger-scale northern gas is difficult since it consists of a complex combination of emission and absorption. However, we can study the absorption structure
on the dynamical centre of the nucleus to find a limit to an inflow velocity. The HCN 3--2 has a shift of +50 \kms\ with respect to $v_{\rm sys}$.  Since we cannot determine exactly how far from the 
nucleus the absorption occurs, we adopt +50 \kms\ as an upper limit to the inflow velocity, $v_{\rm in}$.  

The CS 7--6 absorption extends out to between -150 and -180 \kms\ (with a possible faint outlier out to -280 \kms; Fig.~\ref{f:CS_pV}). It is either the near-face-on outflow velocity we see (in scenario A)  or the near-side of
an inclined outflow in scenario B. In the former case, the outflow velocity is (corrected for $i$=30) $v_{\rm out}=170-200$ \kms, while for scenario B, $v_{\rm out}$ may be significantly higher.  
The escape velocity of the inner region can be set to $v_{\rm esc}$=$3 \times v_{\rm rot}$ \citep{martin05} which is $v_{\rm esc}$=300 - 600 \kms\ and, unless we are observing the outflowing gas edge-on, 
it is unlikely that it is fast enough to escape the centre of IC~860.


\subsection{The nature of the buried source}
\label{s:buried}

The opaque mm, "hot-core" nuclear emission can be fitted with an $r$=8 pc, $N$(H$_2$)=$5 \times 10^{26}$ $\cmmd$ dust structure with temperature $T_{\rm d}$$ \simeq$280 K (see Sect.~\ref{s:dust_opacity}).
Since there may be effects of continuum trapping,   one should be cautious in using the hot-core $T_{\rm d}$ to derive luminosity surface brightness of the inner $r$=8 pc. Assessing the impact of continuum
trapping requires information on source structure and orientation.   A very simple estimate is to consider a 280 K surface radiating against a 100 K surface, then a luminosity of $\sim2 \times 10^{11}$ \lsun\ may be
generated in the inner region. We may also consider the size of the HCN-VIB structure as the edge of the core, since the  HCN-VIB excited by hot dust. Assuming that the HCN-VIB emission is the same as the
hot MIR structure  (see discussion in Sect.~\ref{s:dust_opacity}), the trapping will reduce the luminosity to $10^{11}$ \lsun\   (for $r$=10 pc and $\tau$(MIR)$\sim$5; \citet{kaufman98}). This is a very simple
assumption and investigating how the MIR hot structure relates to the mm one will provide important clues to the properties of the IC~860 CON. For the discussion below, we  assume that $10^{11}$ \lsun\  
is generated in the inner $r$=10 pc.

\subsubsection{Mass budget}

The maximum dynamical mass at $r$=10~pc, $M_{\rm dyn}=9 \times 10^7$  \msun  is for the near face-on ($i$=30$^{\circ}$) disk (see Sect.~\ref{s:hcn-vib_dyn} ). For $L_{\rm IR}\sim 10^{11}$ \lsun, the 
luminosity to mass ratio $L/M$ would be  $\sim10^3$. This is a high value for $L/M$, but it would not exclude star formation as a possible cause of the luminosity at the centre of IC~860, since massive O/B stars
may have $L/M$ approaching $10^4$.  The nucleus of IC~860 will host a SMBH, a pre-existing nuclear stellar cluster (NSC), a potential starburst stellar population  and molecular gas:

\noindent
{\it $M_{\rm NSC}$: }\, Most galaxies have  pre-existing NSCs that exist together with the SMBHs. Since IC~860 has an aging,  post-starburst stellar spectrum, it is likely also to have a relic NSC.
Based on the velocity dispersion we estimate that the NSC has $L_{\rm NSC} \lapprox 10^7$ \lsun\ in the H-band which suggests that  $M_{\rm NSC} \approx 10^7$ \msun\ (see Fig. 7 of \citet{ravindranath01}).

\noindent
{\it $M_{\rm SMBH}$: }\, To estimate the SMBH mass we can use the relation between stellar velocity dispersion, $\sigma$, and that of the SMBH mass, the $M_{\rm SMBH}$-$\sigma$ relation. The SDSS DR9 study lists a stellar velocity
dispersion for IC~860 of $\sigma$=90 \kms\ which implies a SMBH mass of $2 \times 10^6 - 10^7$ \msun  \citep{mcconnell13,graham11}. \citet{mcconnell13} argue that the dispersion is typically calibrated out to the
effective radius. However, the $\sigma$ value may still suffer from extinction and should be viewed with some caution. As an alternative, a relation between the larger-scale disk rotation velocity and $M_{\rm SMBH}$
can be used  \citep{davis19}. Using a value from {\it Hyperleda} on the rotation velocity of IC~860 of $v_{\rm rot}$=226 \kms\  implies a $M_{\rm SMBH}$ of $4 \times 10^7$ \msun. However, we do not know how $v_{\rm rot}$
was determined, so this value should also be viewed with some caution, and we adopt it as an upper limit. 

\noindent
{\it $M_{\rm gas}$: }\, For a disk of $N$(H$_2$)=$5 \times 10^{26}$ $\cmmt$, radius $r$=10 pc and thickness $h$=1 pc, $M_{\rm gas}$=$2 \times 10^8$ \msun. This is higher than the estimated maximum dynamical mass. 
However, if the emitting 1 mm surface is instead at $\tau$=1 and $N$(H$_2$) is $10^{26}$ $\cmmt$, the mass drops to $M_{\rm gas}$=$4 \times 10^7$ \msun\ and we adopt this as a lower limit to the mass. We note also that
some fraction of the nuclear gas may be engaged in in- and outflowing motions, which may not enter into the dynamical mass estimate.

\subsubsection{AGN, starburst, or both?}

The Eddington luminosity of a $4 \times 10^7$ \msun\ SMBH is $2 \times 10^{12}$ \lsun. Thus the luminosity of IC~860 could be consistent with a buried, efficiently (10\% Eddington) accreting SMBH.
Is it also possible that the luminosity could be generated by a starburst within $r$=10 pc?

If we use O5 stars as a proxy, then the mass per star is $\approx$40 \msun\ and $L/M\simeq8000$ \lsun/\msun. We then require $1.5 \times 10^7$ \msun\ of equivalent O5 stars to produce $L=10^{11}$ \lsun.
For a normal Salpeter initial mass function (IMF),  the mass of low-mass stars would be a factor of ten greater than that of the O-stars - requiring a total mass in young stars of $M_{\rm SB}$$\sim 2 \times 10^8$ \msun.
Even if the estimated mass budget above is uncertain, it cannot accommodate the mass of a starburst with a normal IMF. Therefore, to drive the luminosity inside $r$=10 pc with a starburst,
a top-heavy IMF (that produced only massive O-stars) is necessary (and requires tweaking masses for SMBH, NSC, and/or gas down slightly). The stars would need to be formed in less than their lifetimes $t< 3 \times 10^6$ yr, with a high SFR$\simeq$10-20 \msun yr$^{-1}$ purely in massive stars (the steep radio spectrum also appears inconsistent with the primarily thermal pre-supernova emission from young stars). We find that
the luminosity from the inner region would have to be lowered by a factor of ten for it to be powered by star formation with a normal IMF.   \\

If the luminosity of IC~860 is indeed generated inside $r$=10~pc, the IR luminosity surface density would be $\Sigma_{\rm IR} > 10^{14}$ \lsun\ kpc$^{-2}$, typical of Seyfert galaxies \citep{soifer03}. However, \citet{andrews11}
suggest that also hot opaque starbursts may attain very high $\Sigma_{\rm IR}$ that rival those found in AGNs. Perhaps both activities are likely to occur at the same time. In their discussion of the possibility of the existence of  hot 
compact starbursts, \citet{andrews11} remark that "The high surface densities necessary to enter this regime may only be attained in the parsec-scale star formation thought to attend the fueling of bright active galactic nuclei".  
A nuclear ($r<$10 pc)  starburst could require a top-heavy IMF in order to prevent overproduction of low-mass stars. Interestingly, possible enhancements of $^{18}$O have been found in the central
regions of powerful AGNs such as Mrk231 \citep{gonzalez14} and IRAS13120 \citep{sliwa17}. Studying various isotope ratios is a potentially powerful tool to investigate the IMF in a star forming region \citep[e.g.][]{hughes08}. The  $^{18}$O isotope is less abundant than $^{16}$O and is thought to be synthesized by partial He burning in massive stars \citep[e.g.][]{wilson92}. Elevated  $^{18}$O over $^{16}$O ratios are suggested to be an indication of a top-heavy IMF  \citep[e.g.][]{romano17}.   The potential  coexistence of AGNs and starburst activity requires further study.

\section{The evolutionary state of IC~860}

Inflowing molecular gas (with $v_{\rm in} \lapprox$ 50 \kms) is likely responsible for the build-up of exceptional columns of gas and dust in the $r$=9 pc opaque nucleus of the LIRG IC~860,  driving a transient
phase of rapid evolution.  We can link the accumulation of gas and dust in the nucleus of IC~860 to it being a barred, interacting galaxy. The $N$(H$_2$) of IC~860 rivals that of the iconic ULIRG merger Arp~220 
\citep{scoville17,sakamoto17} - albeit possibly on a smaller scale. It is interesting that extraordinary large gas column densities of $N$(H$_2$)=$5 \times 10^{26}$ $\cmmt$ (for standard dust-to-gas ratios) can also be
built up in lower-luminosity systems that are not major mergers.  

The dust enshrouded nuclear activity may be powered by efficient accretion onto a SMBH.  Gas funnelled to the centre then leads to the rapid growth of the SMBH. There is a significant gas reservoir
of $3 \times 10^9$ \msun\ \citep{alatalo16} of molecular gas in the inner region that may be funnelled into the nucleus to fuel the growth;   although it is not clear how this reservoir directly links to the inflow. The
time-scale of gas to reach the nucleus from $r$=50 pc is roughly 1 Myr, and nuclear growth will therefore continue for at least 1 Myr, unless the feedback from the accretion is about to turn off the feeding process (see e.g. \citet{ricci17}).
The inflowing gas may also fuel a coexisting extremely top-heavy nuclear starburst (or a more normal starburst if the gas is deposited further from the nucleus). If so, we are catching IC~860 at a highly specific time
in its evolution where all the high-mass stars have been born but have not yet exploded as supernovae. 

The compact and dense nuclear outflow of IC~860 appears to be behind the foreground, inflowing gas, either because it is very young and compact, or because it is the base of a larger-scale outflow. 
Optical images of IC~860 reveal a large, v-shaped kiloparsec-scale dust structure along the minor axis.  With an outflow velocity of $v_{\rm out}$=170-200 \kms, the nuclear gas is unlikely to escape IC~860,
or be pushed out to kiloparsec-scales, unless it is in the process of being accelerated. Future multi-wavelength studies will reveal if the dense nuclear outflow is the beginning of a continuous outflow that is linked to the optical
dust features, or if the dense outflow represents a very recent outburst in a recurring cycle of nuclear flares,  where the v-shaped, kiloparsec-scale dust lanes are a remnant.

In a recent study of the link between the relative luminosity of HCN-VIB to FIR luminosity ($L_{\rm HCN-VIB}/L_{\rm FIR}$) and the presence of outflows \citep{falstad19}, it is suggested that vibrationally excited HCN traces
a heavily obscured stage of evolution before nuclear feedback mechanisms are fully developed. In this latter study $L_{\rm HCN-VIB}/L_{\rm IR}$ is compared with  outflow and inflows detected in the far-infrared (FIR) through the FIR 119 $\mu$m
as observed by the {\it Herschel} space telescope.  The HCN-VIB luminous galaxies generally show FIR OH inflows, but longer-wavelength high-resolution studies reveal the presence of collimated outflows from an
inclined nuclear dusty disk in CONs such as Arp~220 and Zw~049.057  \citep{varenius16,sakamoto17, barcos18, falstad18}.  IC~860 has a $L_{\rm HCN-VIB}/L_{\rm FIR}$ of $3.2 \times 10^{-8}$ and belongs to the HCN-VIB luminous
galaxy category \citep{aalto15b}.  For IC~860, the presence of a molecular outflow is less striking than in the formerly mentioned CONs (and it appears to be quite slow) while the inflow seems comparatively more prominent.
This points to the possibility that IC~860 is in an extreme phase of its evolution - even compared to other HCN-VIB luminous galaxies. The outflow appears to be in an early stage, and we may be witnessing the onset of feedback
for this cycle of activity.   We note that IC~860 does not have a reported {\it Herschel} OH observation and therefore does not appear in the \citet{falstad19} study of $L_{\rm HCN-VIB}/L_{\rm IR}$ versus FIR OH.

\section{Conclusions}

We used high-resolution (0.\asec 03 to 0.\asec 09 (9 to 26 pc))  ALMA (100 to 350 GHz  ($\lambda$ 3 to 0.8~mm)) and (0.\asec 04 (11 pc) ) VLA 45 GHz  measurements to image continuum and  spectral line emission from the inner (100 pc) region of the nearby IR-luminous galaxy IC~860.   We detect compact ($r<$10 pc (HWHM)), luminous
 3 to 0.8 mm continuum emission in the core of IC~860, with brightness temperatures $T_{\rm B}>160$ K.  The 45 GHz continuum is also compact but significantly fainter in flux than the 3 to 0.8~mm emission.

We suggest that the  3 to 0.8~mm continuum emerges from hot dust with radius $r$=8 pc and temperature $T_{\rm d} \simeq$280 K.  We also suggest that the dust is opaque at mm wavelengths, which implies a large 
 H$_2$ column density of $N$(H$_2$)$\gapprox10^{26}$ $\cmmt$. We assume a standard dust-to-gas ratio of 1/100, and adopt a simple, smooth (non-clumpy) single-temperature geometry.
 There is no indication (based on current information)  of a significant contribution from synchrotron or optically thin or thick, free-free emission to the mm continuum. But more information is necessary to fully assess the
 contribution of free-free emission. 

Vibrationally excited lines of HCN  $\nu_2$=1f J=4--3 and 3--2 (HCN-VIB)  are seen in emission, and are resolved  in the inner 0.\asec 15 (43 pc). The line-to-continuum ratio drops towards the
inner $r$=4 pc, resulting in a ring-like morphology. We propose that this is due to opacity and matching HCN-VIB excitation- and continuum  temperatures.  The emission reveals a north--south nuclear velocity gradient with projected rotation
velocities of $v$=100  \kms\ at $r$=10 pc. The brightest emission  is oriented perpendicular to the velocity gradient, with a peak HCN-VIB 3--2 $T_{\rm B}$ of 115 K (above the continuum).
The enclosed mass inside $r$=10 pc can be estimated to $M_{\rm dyn}=9 \times 10^7$  \msun\ for a disk of inclination $i$=30$^{\circ}$. However, the  disk inclination may be higher ($i>60^{\circ}$) with
an additional east--west outflow component.

Ground-state lines of HCN 3--2 and 4--3, HC$^{15}$N 4--3, HCO$^+$, 3--2 and 4--3 and CS 7--6 show complex line absorption and emission features towards the
dusty nucleus.  HCN and HCO$^+$ have reversed P-Cygni profiles indicating gas inflow with $v_{\rm in} \lapprox 50$ \kms. Foreground, continuum- and self-absorption structures 
outline the flow, and can be traced from the northeast into the nucleus. In contrast, CS and HC$^{15}$N show nuclear blueshifted line profiles with line wings extending out to  -180 \kms. 
We suggest that a dense, compact outflow is hidden behind a foreground layer of inflowing gas and we present two scenarios: one where the disk is near face-on with an additional foreground inflow
component and a compact, young outflow oriented toward us; the other with a near edge-on disk to which the inflow structure connects and where the outflow structure is now oriented largely
away from the observer.

The high opacity in the centre of IC~860 complicates determinations of nuclear luminosity and luminosity density and therefore also aggravates attempts to address the nature of the luminosity source.
Based on a simple estimate, we suggest that $L_{\rm IR}\sim 10^{11}$ \lsun\ is emerging from the inner 10 pc.  The luminosity may be generated by an AGN and/or a starburst. If a significant fraction ($>$10\%)
of the nuclear luminosity is emerging from star formation, a top-heavy IMF is required. Regardless of power source, the nucleus of IC~860 is in a phase of rapid evolution where an inflow is building up a
massive nuclear column density of gas and dust.  This gas feeds the central star formation and/or AGN activity of IC~860. The slow outflow appears to be in an early stage, and we may be witnessing the onset of feedback for this cycle of activity.

\begin{acknowledgements}
This paper makes use of the following ALMA data: ADS/JAO.ALMA\#2015.1.00823.S and 2016.1.00800.S.
ALMA is a partnership of ESO (representing its member states), NSF (USA) and NINS (Japan),
  together with NRC (Canada), MOST and ASIAA (Taiwan), and KASI (Republic of Korea), in
cooperation with the Republic of Chile. The Joint ALMA Observatory is operated by
ESO, AUI/NRAO and NAOJ.
We acknowledge excellent support from the Nordic ALMA Regional Centre (ARC) node based at Onsala Space Observatory. 
The Nordic ARC node is funded through Swedish Research Council grant No 2017-00648.
SA acknowledges that this project has received funding from  the European Research Council (ERC) under the European Union's 
Horizon 2020 research and innovation programme, grant agreement No ERC-2017-ADG-789410. SA also acknowledges the
Swedish Research Council grant 621-2011-4143. KS was supported by grant MOST 102-2119-M-001-011-MY3
SGB thanks support from Spanish grant AYA2012-32295. We thank E. Gonzalez-Alfonso for alerting us to the potential
impact of photon trapping on $T_{\rm d}$.

\end{acknowledgements}

\bibliographystyle{aa}
\bibliography{vib_ref}

\end{document}